\documentclass[11pt,a4paper] {article}
\usepackage{amssymb,amsmath,amsfonts}
\usepackage[dvipsnames]{xcolor}
\usepackage{graphicx}
\usepackage{longtable}
\usepackage{verbatim}
\usepackage{color}
\usepackage[mathscr]{euscript}
\usepackage{framed}
\usepackage{mdframed}
\usepackage{amsmath}
\usepackage{amsfonts}
\usepackage{mathrsfs}
\usepackage{amssymb}
\usepackage{mathrsfs}  
\usepackage{subcaption}
\usepackage{caption}

\usepackage[left=26mm,top=20mm,right=26mm,bottom=20mm]{geometry}

\usepackage{diagbox}

\usepackage{color}
\usepackage{cite}
\usepackage{hyperref}
\hypersetup{colorlinks, citecolor=bluscuro, linkcolor=black, urlcolor=bluscuro}
\definecolor{rossos}{cmyk}{0,1,1,0.55}
\definecolor{bluscuro}{rgb}{0.15, 0.2, .85}
\definecolor{bluchiaro}{cmyk}{1,.3,0.,0.1}

\graphicspath{{./Figures/}}

\newcommand{\be}{\begin{equation}}
\newcommand{\ee}{\end{equation}}
\newcommand{\bea}{\begin{eqnarray}}
\newcommand{\eea}{\end{eqnarray}}
\newcommand{\beq}{\begin{equation}}
\newcommand{\eeq}{\end{equation}}

\def\beqa{\begin{eqnarray}}

\def\eeqa{\end{eqnarray}}

\def\lsim{\mathrel{\rlap{\lower4pt\hbox{\hskip0.5pt$\sim$}}
    \raise1pt\hbox{$<$}}}         
\def\gsim{\mathrel{\rlap{\lower4pt\hbox{\hskip0.5pt$\sim$}}
    \raise1pt\hbox{$>$}}}         

\usepackage[normalem]{ulem}

\newcommand{\arXiv}[2]{\href{http://arxiv.org/pdf/#1}{{\tt [#2/#1]}}}

\makeatletter
\renewcommand{\citation}[1]{%
  \g@addto@macro{\citation@list}{,#1}%
}
\newcommand*{\citation@list}{} 
\newcommand{\sortbibitem}[2]{%
  \global\@namedef{bibitem@#1}{%
    \bibitem{#1} #2
  }%
}
\newcommand{\sort@bibitems}{%
  \@for\next:=\citation@list\do{%
    \@nameuse{bibitem@\next}%
    \global\@namedef{bibitem@\next}{}%
  }%
}
\expandafter\def\expandafter\endthebibliography\expandafter{%
  \expandafter\sort@bibitems\endthebibliography
}
\makeatother

\begin{document}

\vspace{0.1in}

\begin{center}
{\Large\bf\color{black}
 Gravitational Waves from  Collapse of Pressureless Matter in the Early Universe
}\\
\bigskip\color{black}
\vspace{.5cm}
{Ioannis Dalianis{\large${}^a$}
\vspace{0.3cm} and \vspace{0.3cm} 
Chris Kouvaris${}^b$} \\[5mm]
{
\it {$^a$\,Department of Physics, University of Cyprus, \\
Nicosia 1678, Cyprus} \\
\it {$^b$\,Physics Division, National Technical University of Athens\\ 15780 Zografou Campus, Athens, Greece
}}\\[2mm]

\end{center}

\vskip.2in

\noindent
\rule{16.6cm}{0.4pt}

\vspace{.3cm}
\noindent
{\bf \large {Abstract}}
\vskip.15in 
If an early matter phase of the Universe existed after inflation with the proper power spectrum, enhanced density perturbations can decouple from the Hubble flow, turn around and collapse. In contrast to
what happens in a radiation dominated Universe where pressure nullifies deviations from sphericity in these perturbations, in a matter dominated Universe, the lack of pressure although on the one hand facilitates the gravitational collapse, it allows small deviations from sphericity to grow substantially as the collapse takes place. The subsequent collapse is complicated: initially as non-spherical deviations grow, the collapsing cloud  takes the form of a ``Zel'dovich pancake". After that, the more chaotic and nonlinear stage of violent relaxation begins where shells of the cloud cross and the matter is redistributed within a factor of a few of the free fall timescale, reaching a spherical virialized state. During the whole process, strong gravitational waves are emitted due to the anisotropy of the collapse and 
the small time interval that the effect takes place. The emission of gravitational waves during the stage of the violent relaxation cannot be easily estimated with an analytical model. We perform an $N$-body simulation to capture the behaviour of matter during this stage in order to estimate the precise spectrum of gravitational waves produced in this scenario.

\noindent

\bigskip

\noindent
\rule{16.6cm}{0.4pt}

\vskip.4in

\section{Introduction}

Strong gravitational waves (GWs) can be emitted during a gravitational collapse if the collapse involves aspherical 
and rapid  motion of dense matter distributions \cite{Misner:1973prb}.
 Astrophysical systems are common examples of strong gravitational sources.
Presumably, such conditions are also  found  in the dense and energetic early Universe if 
there are sufficiently large density perturbations.
Primordial density perturbations are observed at CMB scales through temperature anisotropies and have a minute size.
 At small scales the amplitude of density perturbations is not measured and it might be significantly larger \cite{Carr:2020gox, Carr:2017edp, Dalianis:2018ymb, Gow:2020bzo, Yang:2020egn, Ando:2022tpj}. 
  Deviations from homogeneity in the primordial Universe can result in gravitational instabilities and  perturbative 
GW emission \cite{Matarrese:1997ay, Mollerach:2003nq}. 
Furthermore, if an  early matter domination era (EMD) has been realized, the 
density perturbations grow further and in the absence of background pressure an isotropic evolution 
 cannot be supported. 
As a result,   an additional strong GW component is generated during the process of matter compression and  virialization, leading possibly to the formation of ephemeral halos. GW emission of this sort, that involves 
pressureless bulk mass motion, 
 is not expected for smooth unclustered energy components, like radiation, that is understood to collapse and expand isotropically.

During an EMD  era  the pressure is vanishing and deviations from sphericity are assumed 
for the primordial density perturbations that enter the horizon. Initially in the  linear regime,  the  perturbations evolve in the expanding background and asphericity  grows.
At some point the expansion of the overdensities is expected to halt along the dimension that the compression is maximal and collapse follows.
The first stage of  gravitational evolution 
can be described by the ``pancake formation"  process, developed by Zel'dovich \cite{Zel}. 
The Zel'dovich approximation is a theory for the growth
and  evolution of  density cosmological  perturbations in the nonlinear stage and in a background of zero pressure. 
Central quantities are a deformation tensor $D_{ij}$, that describes the 
non-spherical  geometrical profiles of density perturbations, 
and a probability density function (PDF) associated with the 
statistical significance of the
aspherical  configurations. 
Assuming an initial  Gaussian distribution for the perturbations,
probabilities of the various initial 
profiles
 can best be obtained using the 
  Doroshkevich PDF  \cite{Doroskevich1970}.

We will consider and study the evolution of a gravitational instability 
originated from strong linear perturbations 
 in an expanding EMD  Universe. Our main focus is on the gravitational radiation emitted during
  the anisotropic motion of matter which deviates from the background Hubble flow.  
   The spectrum of the GWs produced in the first stages,  
   during which  the inhomogeneity decouples 
   from  the background expansion and experiences a pressureless collapse, 
   has been  computed  in ref. \cite{Dalianis:2020gup}.
In that first paper,  the analytical expression for the spectrum of stochastic GW background has been given in  the post-Newtonian framework  which describes the GW emission during the entire course  of the evolution of  the inhomogeneity. 
The computation of the spectrum was done  assuming the Zel'dovich solution 
that is valid roughly until  the limiting stage of the pancake collapse. 
 This limitation  on the description of the nonlinear evolution 
 calls for more sophisticated numerical approaches 
  and motivates the current work. 
In this paper we take a step forward  and complement our previous results  \cite{Dalianis:2020gup}
 by tracking the evolution of the inhomogeneities deep in the nonlinear regime 
  via  numerical means.
We utilize  $N$-body techniques   which provide a unique theoretical tool that opens up the way to go beyond the analytic approximations. 
The underlying assumption  made  is that the fluid which realizes the  EMD era  can be simulated  by a distribution of individual nonrelativistic particles. 
Via $N$-body simulations  we find that the dynamical evolution of the perturbation into nonlinear structures 
 involves 
 bulk mass motions that source GW emission, occasionally with pretty large amplitude.
We compute the quadrupole for the mass distribution and obtain the spectrum of GWs  
 from the moment of horizon entry of the perturbation  until the  stage of nonlinear  clustering.

 We use
 the publicly available Nbody2 code of  Aarseth  \cite{Aarseth:2001zp}, which is a Newtonian gravity $N$-body code, to run our simulations.
The spectrum pattern of    GWs emitted  by subhorizon modes
   is found to 
 depend significantly on the    
  initial conditions that characterize the geometry of the density perturbations. 
The spatial structure of the initial  density 
and velocity 
perturbations 
 are specified by the deviation of the density perturbations  $\sigma$ and the  parameters
  $(\alpha, \beta, \gamma)$ 
 of  the deformation tensor $D_{ij}$. 
 These 
 parameters define a particular configuration and determine our initial conditions for 
 the 
  $N$-body simulations.  
  We assume small-scale perturbations with wavelength $k^{-1} \ll 1\, {\rm Mpc}^{-1}$ and a large deviation  $ 0.005 \lesssim \sigma \lesssim  0.2$, so as the processes involved to have  potentially observable implications. 
We track and describe the  cosmological collapsing processes 
 running the $N$-body system for sufficiently many crossing times.
We observe that 
the collapsing matter undergoes violent relaxation  
and settles into an  equilibrium configuration.
After the completion of the phase mixing and relaxation process  we observe that a  halo forms which satisfies the virial theorem, i.e.  virialization takes place.

 The gravitational emissivity is found to be very sensitive to the geometrical features, determined by the deformation parameters $(\alpha, \beta, \gamma)$ and $\sigma$ of the density  perturbation.
 A Hubble patch that encloses a perturbation with an acute 
 nonspherical profile emit GWs with a power several orders of magnitude larger than a patch with an approximate spherical profile.
The probability for each profile to be realized also varies significantly and  is determined by the Doroshkevich PDF, so that  profiles with extreme parameters are extremely rare.
The average GW spectrum is found after  a summation procedure, that  takes into account the statistical significance of each configuration. 
The observable average spectrum 
is  cosmology  dependent.  Aside from the $\sigma$ value, it is specified after values for the mass of the perturbation $M$ and  the reheating temperature   
 $T_{\rm rh}$ are given.
In particular, the overall frequency is specified  by  the  mass $M$ parameter, which is associated with   the wavenumber position $k$ of the curvature spectrum peak ${\cal P_R}(k)$.
Here we will consider curvature power spectra peaks that are dominantly monochromatic  with amplitude $\sim \sigma^2$.
Both  amplitude and frequency have also an overall dependence  
on the duration of the EMD phase parameterized by the  $T_{\rm rh}$ value.

We run the  $N$-body simulation with initial positions and velocities for the particle distribution specified 
at the moment of maximum expansion of the inhomogeneity $t_{\rm max}$ which is the moment in time that the first turnaround occurs, set to be along the axis $r_1$. 
We adopt  the Zel'dovich solution, which we consider reliable enough until the time of the turnaround.  Zel'dovich solution for determining initial positions and velocities have been also used elsewhere \cite{Szalay:1984df}.
However, we observe that the accuracy of the Zel'dovich  description is gradually lost until  the stage of the pancake collapse $t_{\rm col}$ and eventually breaks down.
As expected the spectrum of the produced  GWs found from the $N$-body results
has similarities to the spectrum found using the Zel'dovich solution   \cite{Dalianis:2020gup} only for times before the pancake collapse and it is characterized by a rough single peak associated with the undergoing quadrupole changes.

Going beyond  the stage of pancake collapse, the $N$-body simulation 
reveals a new dynamical evolution. The distribution of  $N$-particles experiences a violent relaxation that drives it, sooner or later, towards a quisi-steady spherical configuration. This stage produces GWs that might or might not surpass the amplitude produced during the first stage of pancake collapse.
In particular, initial configurations with a  mostly spherical shape reach the virialized state in a time  $t_\text{vir}$ roughly twice the time of pancake collapse without  increasing the amplitude of the final GW spectrum much.  Configurations with acute asymmetry  experience a second  collapse after the pancake stage.  Fast rearrangements of the mass distribution result in an enhanced amplitude for the GW spectrum and produce additional spectral peaks.  The maximal enhancement is found to reach one order of magnitude in size.
The notable features
are clearly observed in configurations that experience an oblate-prolate geometrical change during the course of virialization. These geometrical changes generate a second rough peak in the GW spectrum at a larger frequency with a value associated with the  duration of the violent relaxation process. 
This second peak gets modulated or even disappears  as we change the values for the deformation parameters $\alpha, \beta$ and $\gamma$.
 The emission of GWs 
 practically shuts down after the system of particles settles in the virialized state. The time required to achieve virialization  also varies for different  configurations with shorter values for $t_\text{vir}$ resulting in stronger GW emission, something that is expected since smaller timescales mean larger derivatives of the quadrupole moment and therefore increased GW signal.

The total GW signal is an average found only after a sufficiently large  number $N_\text{conf}$ of different configurations is properly considered. 
We  apply a systematic discretization process to scan the continuous spectrum of the parameter space. 
This  requires
the repetition of the numerical experiment 
about $N_\text{conf}\sim 5000$ times, each time with different initial conditions.  
This way we test and record the evolution of $N_\text{conf}$ initial  asphericities  dictated by the deformation parameters 
 $(\alpha, \beta ,\gamma)$
and with a common variance $\sigma$.  
The implementation of such a large number of $N$-body runs 
has been made possible by the use of a small computer-cluster.
The computational  limitations restricted us to consider a relatively small, though seemingly sufficient number of particles  $N\sim 10^4$.
The total average  GW signal obtained can be viewed as a nonlinear superposition of GWs emitted by an assembly of $N_\text{conf}$ patches each enclosing a perturbation with statistical significance 
dictated by the Doroshkevich $F_D$ distribution.
This averaging procedure yields a GW spectrum characterized by new interesting features.
Instead of a single smooth peak, which is  given by the Zel'dovich solution, there is a  two-peak structure.
The extra peak is attributed to the violent relaxation processes that 
are not captured by the Zel'dovich framework.
Actually,
hump-like  features around the peak appear at the GW spectrum already before the average time of pancake collapse. 
This is a fine-structure difference compared to the Zel'dovich  result 
and it  is understood from the fact that shell-crossing processes do not occur simultaneously in all the configurations.
The  recognition  that there is a specific GW signature of the virialization process on the GW spectrum is one of the main results of this work.

Our primary interest here is for gravitational collapse processes taking place
on scales that enter the Hubble horizon before the big bang nucleosynthesis (BBN), 
though our results can be applied to processes occurring after recombination as well. 
The 
assumption of an EMD era realized before BBN is not standard, however it is common  and  natural in several Beyond the Standard Model (BSM)  set-ups.
Actually, the earliest epoch we can confidently gain information about
the early Universe is for temperatures ${\cal O}(1)$ MeV where  BBN takes place \cite{Allahverdi:2020bys}. In this aspect GWs account for a unique and very valuable probe, see e.g. \cite{Roshan:2024qnv} for a recent review.
For larger  energy scales  
we are ignorant about the energy content of the Universe and only 
at ultra high energy scales an inflationary stage is postulated. 
Between the BBN epoch and the Planck or the inflationary era stretches an extensive,  in terms of e-folds, cosmic period  
where a deviation from the  steady thermal state (radiation domination (RD)) cannot be 
excluded.
Common examples that can realize an EMD  are the slow decay of the inflaton field or the slow decay of massive metastable particles that dominated the total energy density.
What is of particular interest is that these fields can naturally realize an EMD era if they act as scalar condensates.
 Hence, the virialized halo that we observe  in our simulations is  an ephemeral one that we assume to finally  evaporate due to particle decays that reheat the Universe. We will restrict ourselves to an idealized pure EMD era and we will not explore the possible implications of the gradual decay of the condensate.

  GWs production during EMD at linear order cosmological perturbation theory has been studied in refs. \cite{Inomata:2019ivs, Inomata:2019zqy}.  
 At second order in the expansion of cosmological perturbations, tensor perturbations are sourced, dominantly, by quadratic combinations of first-order scalar perturbations \cite{Ananda:2006af, Baumann:2007zm, Malik:2008im}.
  The GW amplitude is found to have a strong sensitivity on  how fast or slow  the transition from EMD to RD is \cite{Pearce:2023kxp}.  GWs are also sourced by possible isocurvature perturbations of the fluid component that realizes the EMD, see e.g.  \cite{Papanikolaou:2020qtd, Domenech:2020ssp, Domenech:2023jve, Inomata:2020lmk} for GWs from the primordial black hole (PBH) dominated early Universe.
An extra motivation to study  GW production during EMD is given by several proposals of nonlinear structure formation which is not necessarily washed out by the reheating \cite{Padilla:2021zgm, Yoo:2021fxs, Durrer:2022cja, Flores:2022uzt, DeLuca:2022bjs, Domenech:2023afs,  Flores:2023dgp}.
The limitations of the perturbative analysis  due to the presence of a nonlinear scale \cite{Assadullahi:2009nf}  restricts the applicability of the existing formalism which describes the production of secondary  GW during EMD. 
In this respect, our analysis can be viewed as complementary to the perturbative one  adding further insights into particular aspects of the nonlinear processes involved in the  gravitational collapse.
 Earlier studies \cite{Jedamzik:2010hq, Nakama:2020kdc, Eggemeier:2022gyo}  have explored the GW production during an EMD era and within the non-perturbative regime $N$-body  numerical tools were utilized.
More recently ref. \cite{Fernandez:2023ddy} performed  $N$-body simulations taking into consideration the decay of  the matter fluid incorporating  the impact of reheating. 
In our work we give a comparative study between semi-analytic Zel' docivh and the corresponding numerical $N$-body results and  highlight quantitatively the implications of the virialization stage on the GW spectrum. 

The second  assumption we make is that the amplitude of the power spectrum of primordial curvature perturbations rises at small scales (although our results could be adjusted to the case of a scale invariant spectrum). The minimal assumption  is that  the spectrum of primordial fluctuations ${\cal P_R}(k)$ is nearly flat over a wide range of length scales. 
Deviations from flatness could have arisen naturally on some small scale if there is a strong feature in the inflaton potential, see e.g. ref. \cite{Dalianis:2023pur} for a recent review.
The associated length scale may be far below the scales observed in  cosmic structures and correspond to small or even tiny horizon masses. 
Examples of overdensities with masses of order of magnitude $10^{-21} M_{\odot}$, $ 10^{-12} M_{\odot}$  and $10 M_{\odot}$ will be explicitly discussed. 
If the power of such a  high-wavenumber fluctuation is significant a detectable GW signal can be produced. It is actually exciting that at low frequency bands PTA experiments report a stochastic GW background signal. This signal  has already implications on early Universe cosmological scenarios, see e.g. \cite{Balaji:2023ehk, Zhao:2023joc, Chang:2023aba, Domenech:2024rks}. 
In higher frequency bands,  LIGO-Virgo-KAGRA are putting constraints on the amplitude of the background.
Additionally to the stochastic GW background, a PBH population in a mass range analogous to that of the overdensity  might be produced during EMD \cite{Khlopov:1980mg, Polnarev:1985btg, Harada:2016mhb, Harada:2017fjm, Dalianis:2018frf, Kokubu:2018fxy, Ballesteros:2019hus, DeLuca:2021pls} and contribute to the dark matter in the Universe. The correlation between the PBH abundance  and the associated GW spectrum,  see e.g. \cite{Garcia-Bellido:2017aan, Dalianis:2020cla, Fumagalli:2021cel, Dalianis:2021iig, Yuan:2021qgz, Domenech:2021ztg, Chen:2022dah,  Balaji:2022dbi, Meng:2022low, Boutivas:2022qtl, Choudhury:2023fwk, Franciolini:2023pbf, Yuan:2023ofl, Ianniccari:2024bkh, Domenech:2024cjn} for some recent  discussion, is a rather fascinating aspect of this cosmological scenario.

The structure of the paper is as follows. In sec. \ref{sec: set-up}, after a brief introduction to the Zel'dovich approximation, 
we overview the salient aspects of the $N$-body  numerical techniques.  We describe  the initial conditions for the density perturbations and the methodology  that we follow to run our simulations. 
In this respect we quote the  Doroshkevich PDF which specifies the statistical significance of the initial profile for the shape-fluctuations considered.
In sec. \ref{sec: GWs}
we describe the formalism for the computation of the GW spectrum within the quadrupole approximation.   We present in detail the dynamics of the $N$-body distribution and  the subtle  evolution of the density profile identifying the critical stages.
We compute the gravitational radiation emitted during the numerically simulated collapse and we compare it with the previous results obtained within the Zel'dovich (semi)-analytic framework. 
In sec. \ref{sec: average}
we describe the probability sampling method  we follow to compute the overall average 
 GW signal. We discuss particular examples of GWs signals that can be potentially detectable by operating or future GW detectors, giving also emphasis to the recent positive results of PTA experiments.  We finish in sec. \ref{sec: concl} with a discussion and our conclusions.


\section{ The theoretical set up} \label{sec: set-up}

In an earlier study \cite{Dalianis:2020gup} we followed the Zel'dovich approximation to describe the nonlinear evolution of the density perturbation and compute the emitted gravitational radiation until the stage of pancake collapse.
Our goal in this work is to complement this earlier study by investigating the evolution of the triaxial shapes of various  simulated systems until they reach the virialized  state and compute the associated GW emission.

We assume that at some cosmic  moment $t_k$ a linear density perturbation mode with mass $M$ enters the cosmological horizon. 
Initially the Hubble flow stretches the linear inhomogeneity with the spatial size growing proportional to the scale factor. 
 The relative amplitude of the linear perturbation grows with its energy density
decaying slightly more slowly than that of the background. 
When the perturbation amplitude reaches unity
 the linear description is inadequate and a nonlinear description should takeover.
 At the time labeled $t_\text{max}$ the gravitational field created by the perturbation leads to a contraction which overwhelms the Hubble expansion. In the spherical limit  this time corresponds to 
 maximal spatial size of the overdensity.
For  nonspherical 
configurations it is the moment of the turnaround of one out of three spatial directions.

\subsection{Zel'dovich approximation}
For pressureless matter the geometrical shapes of realistic inhomogeneities are typically far from spherical and their collapse is strongly anisotropic. At a first approximation the anisotropic collapse 
can be described by the Zel'dovich solution.
 This solution describes the nonlinear behavior of a perturbation with arbitrary shape, superimposed on three-dimensional Hubble flow. 
 The relation between the Eulerian $r_i$ and Lagrangian $q_i$ coordinates can be written as \cite{Mukhanov:2005sc}
\begin{equation}
r_i=a(t)q_i +b(t)p_i(q_j),
\label{coord}
\end{equation}
where $a(t)$ is the usual scale factor encoding the expansion of the Universe, 
 $b(t)$ is a  growing mode associated with the gravitational instability in a pressureless EMD Universe and $p_i$ are deviation vectors that depend on the shape of the initial perturbation. 
The motion of a group of particles around $q_i$ is described via a deformation (strain) tensor $D_{i j}  ={\partial r_i}/{\partial q_j}$. 
 After choosing a comoving coordinate system where the matrix $\partial p_i/\partial q_j$ is diagonal 
\begin{equation}
\frac{\partial p_i}{\partial q_j}=-\text{diag}(\alpha, \beta, \gamma) \,,
\end{equation} 
it is $D_{i j} =\text{diag}(a-\alpha b, a-\beta b, a-\gamma b)$.
All the properties of the deformation tensor are governed by the density-perturbation spectrum.

A  perturbation characterized by the size $q=k^{-1}$  enters the horizon at $t_k$ satisfying $a(t_k)q=H_k^{-1}$, where $H_k \equiv H(t_k)$ is the Hubble scale at the time $t_k$. 
For simplicity we consider monochromatic  perturbation.
One can relate the horizon entry time $t_k$ with the mass $M$ contained
 \begin{equation}
 t_k=\frac{4GM}{3}.
 \end{equation}
We assume initially small deviations from sphericity, $\alpha b(t_k)/a(t_k) \ll1$, $\beta b(t_k)/a(t_k) \ll1$, $\gamma b(t_k)/a(t_k) \ll1$ so that  at $t_k$ the 
triaxial perturbation is nearly inside the Hubble sphere, $r_i(t_k)\approx  a(t_k)  q$, where $\vec{r}$ is the 
position of the ellipsoid boundary.
At the moment of horizon entry
the density contrast field at linear order is  
\begin{equation} 
 \delta_{\rm L} (t_k) =(\alpha+\beta+\gamma) \frac{b(t_k)}{a(t_k)} \ll 1 ~.
 \label{ratio1}
\end{equation}
During matter domination the density perturbation in the linear regime grows as  $\delta \propto a$ and hence  $b \propto a^2$.
The deformation tensor changes much more rapidly than the Hubble flow factor $a(t)$ and according to  eq. (\ref{coord}) a turnaround will occur  along one of the three axes. The turnaround direction is identified with  $r_1$ after assuming the hierarchy for the deformation parameters
\begin{equation} \label{hiera}
\alpha>0\, , \quad -\infty<\gamma \leq \beta \leq \alpha<\infty  \quad \text{and}\quad \alpha+\beta+\gamma >0 \,.
\end{equation}
 Four important moments are identified:
the horizon entry $t_k$, the maximum expansion time $t_\text{max}$,  the collapse time $t_\text{col}$ and, for special initial conditions, the black hole formation time $t_\text{BH}$. 
An extra important moment $t_{\rm virial}$, that signifies the virialization stage,
  will be introduced in the next subsection. 
At the moment of maximum expansion 
 the mass is about to shrink,  $\dot{r}_1(t_\text{max})=0$ 
with the primordial perturbation along $r_1$  having  half of the size   compared to the unperturbed expansion, $r_1(t_\text{max})= a(t_\text{max})q_1/2$.
This is the turnaround radius $r_1(t_\text{max})$ at which the expansion halts in this direction and collapse commences.
In configurations with spherical symmetry all shells are assumed to have zero velocities at the same time.
For non-spherical configurations the turn-around radius is different in every direction.
During matter domination it is $a(t) \propto t^{2/3}$  and the ratio $a(t_{\text{max}})/{a(t_k)}$
  yields the following expression for the moment of maximum expansion, 
\begin{equation} \label{tmax}
t_\text{max}=\left( \frac{\alpha+\beta+\gamma}{2\alpha \delta_{\rm L}(t_k)} \right)^{3/2} t_k \,.      
\end{equation}
Using Eq.~(\ref{coord}) and the fact that $b(t)\propto  a^2(t)$,  we can write within the analytical approximation (Zel) the evolution of the principal semi-axis of the triaxial ellipsoid  
as a function of time,
\begin{align} \label{ri}
r_i^{\rm Zel}(t, \alpha, \beta, \gamma, \delta_{\rm L})  =\frac{3}{2}t_k^{1/3} t^{2/3}\left(1- \frac{\xi_i}{2}\left(\frac{t}{t_\text{max}}\right)^{2/3}\right)\quad\quad \quad {\rm for } \quad t<t_\text{col}  \,,
\end{align}
where $\xi_i$ takes the values $\xi_i=\{1, \beta/\alpha, \gamma/
\alpha \}$ for the three directions respectively and $t_\text{max}=t_\text{max}(\alpha, \beta, \gamma, \delta_{\rm L})$. 
For times $t>t_\text{max}$ there will be an infalling flux of  matter
and the perturbation will collapse first in one dimension  forming a two dimensional sheet, called Zel'dovich pancakes.  
The moment that $r_1(t_\text{col})=0$ it is
\begin{equation}
t_\text{col}=2\sqrt{2} \, t_\text{max}\,.
\end{equation}
A normalization for the growing mode is 
\begin{equation} \label{norm}
\frac{b(t)}{a(t)}=\frac{a(t)}{a(t_k)}
\end{equation}
With this normalization choice the density field is $\delta_{\rm L}=\alpha+\beta+\gamma$ and $t_{\rm max}=(1/2\alpha)^{3/2}t_k$.
 Also, the perturbation  is a triaxial configuration
with axes ratios 
$1-\alpha :1-\beta : 1-\gamma$.

From the expressions (\ref{ri}) we see that, within the Zel'dovich approximation, the principal semi-axes of the ellipsoid contract only for $\xi_i >0$. By definition  
 contraction happens automatically for the $r_1$ direction because $\alpha$ is positive definite.  The other two directions can either collapse, for positive $\beta$, $\gamma$  or  expand  for negative $\beta$, $\gamma$.  
It is however  expected that  after the formation of the Zel'dovich  pancakes  the other directions will eventually collapse 
ending up to a gravitationally compact structure, 
 where the overdensity  has reached a certain level. 
Apparently, the Zel'dovich description gradually fails to track the evolution of the gravitational collapse and eventually breaks down at the stage of the pancake.

\subsection{$N$-body  description}

\begin{figure}[t!]
  \begin{subfigure}{.322\textwidth}
  \centering
  \includegraphics[width=1 \linewidth]{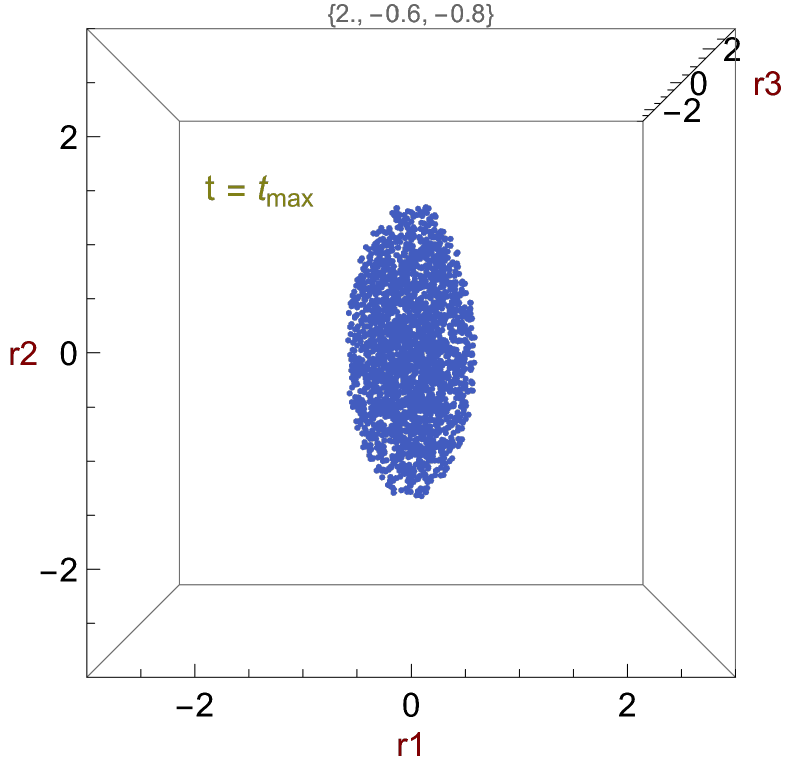}
\end{subfigure} 
  \begin{subfigure}{.322\textwidth}
  \centering
  \includegraphics[width=1\linewidth]{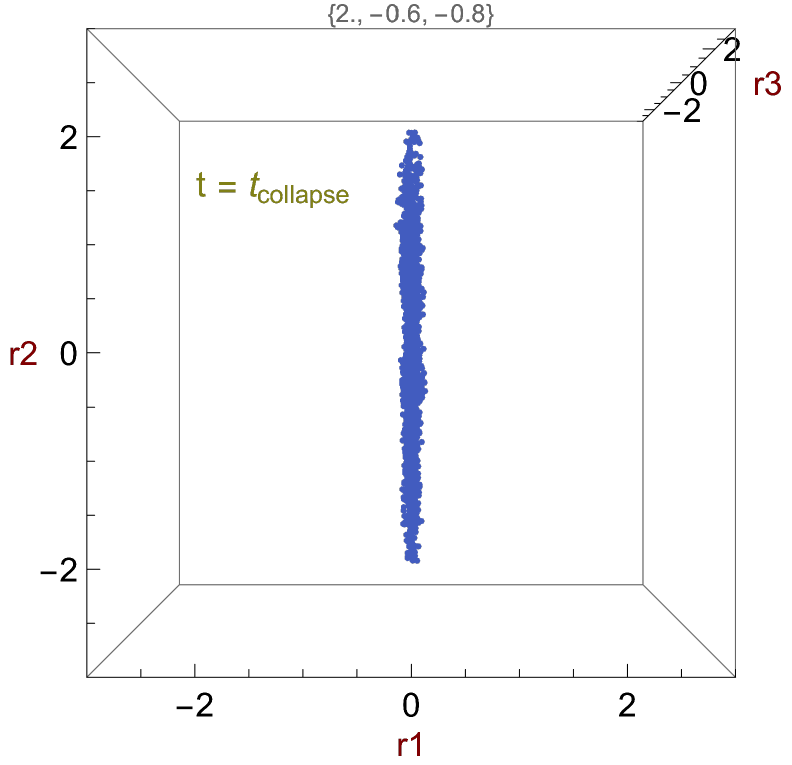}
\end{subfigure}
\begin{subfigure}{.322\textwidth}
  \centering
  \includegraphics[width=1\linewidth]{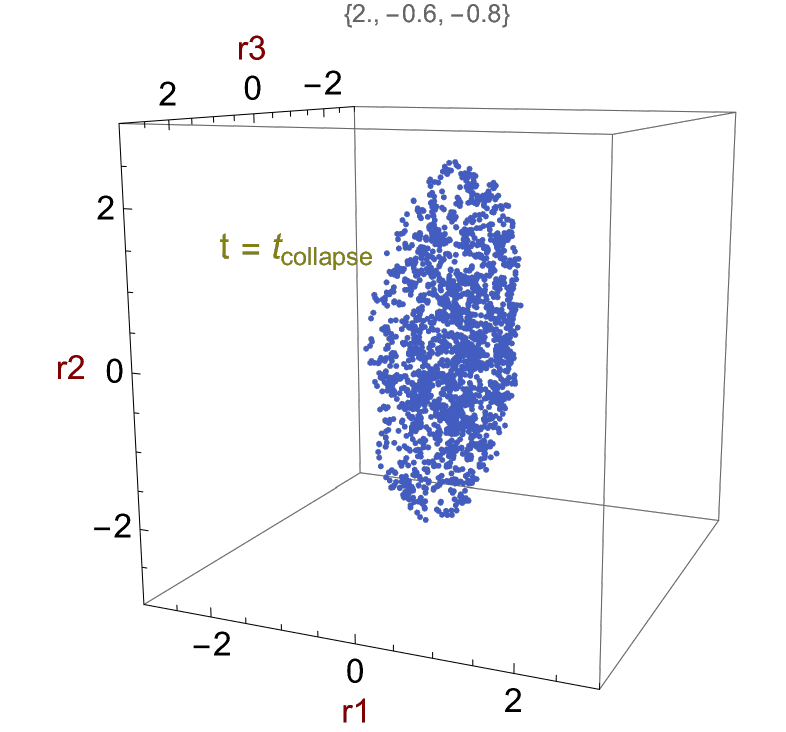}
\end{subfigure}
\begin{subfigure}{.322\textwidth}
  \centering
  \includegraphics[width=1\linewidth]{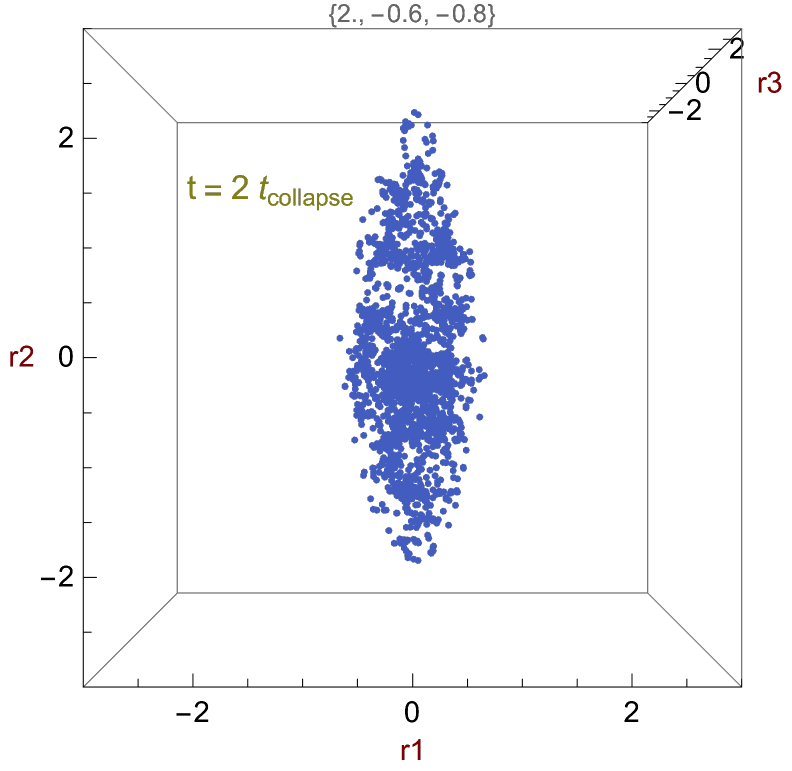}
\end{subfigure}
\begin{subfigure}{.322\textwidth}
  \centering
  \includegraphics[width=1 \linewidth]{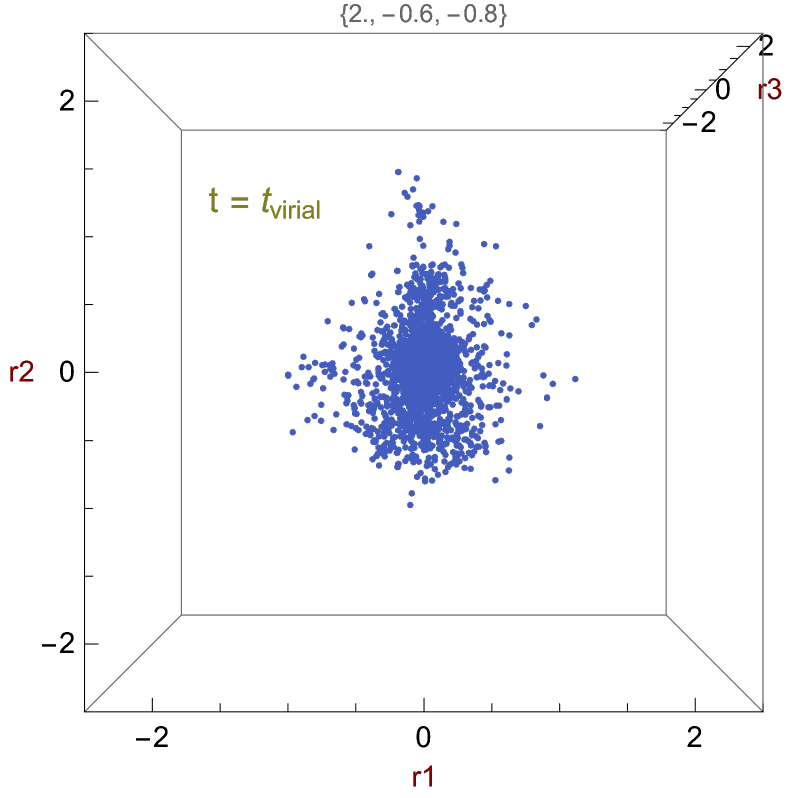}
\end{subfigure} 
  \begin{subfigure}{.322\textwidth}
  \centering
  \includegraphics[width=1\linewidth]{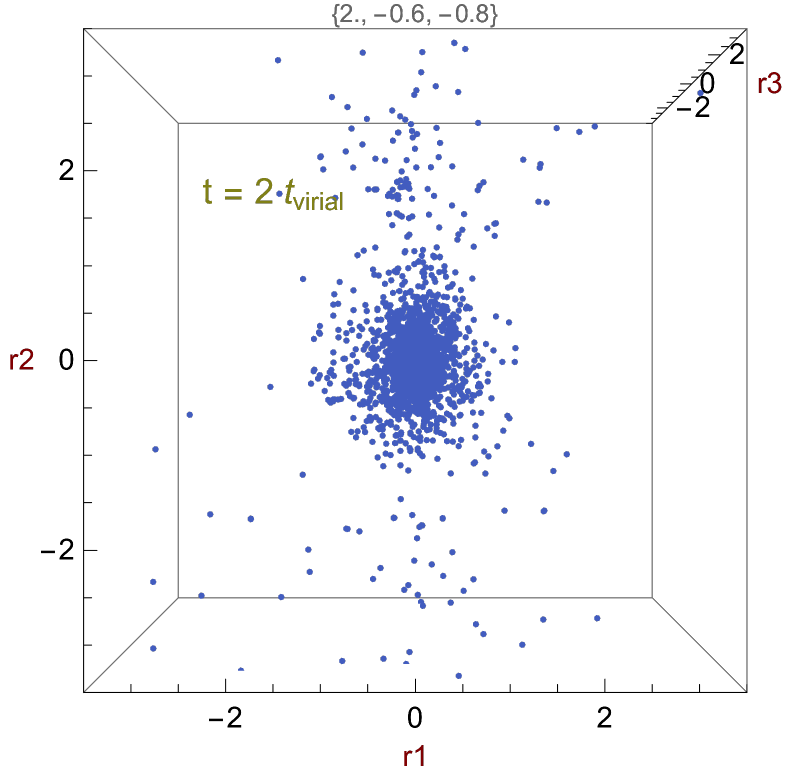}
\end{subfigure}
 \caption{\label{fanim}~~  
The panels  show a projection of the orbits of a cloud of $N=2500$  particles  on the $r_1$-$r_2$ plane for a density perturbation with $\sigma=\sigma_3\sqrt{5}=0.1$ and deformation parameters $(\alpha/\sigma_3, \beta/\sigma_3, \gamma/\sigma_3)=(2, -0.6, -0.8)$.
  }
\end{figure}

The lack of an accurate  description of the collapsing process via simple analytic means   
necessitates the use of  $N$-body numerical tools.   
The density perturbation can be described by a collection 
of mass particles with positions and velocities respectively 
\begin{equation}
\vec{x}_i(t) \,,\quad\quad  \vec{v}_i(t)\,, \quad \quad i=1,... ,N
\end{equation}
 where the index $i$ runs over the particles of the distribution. 
In our set up the dynamics over timescales $t\ll t_\text{reh}$ is that of a collisionless system in which the constituent particles move under the influence of the gravitational field generated by 
the time-varying mass distribution.
$N$-body simulations produce a numerical solution   for arbitrary large times 
and reveal the subtle details of the gravitational collapse.
In fig. \ref{fanim} an animation of the evolution of the particle distribution is displayed in snapshots at the moments of maximum expansion, pancake collapse and finally when the system has settled to approximate equilibrium, called virialization. The moment $t_{\rm virial}$ indicates the initiation of the virialization stage. This moment is identified as a bottleneck stage  associated with the evolution of the quadrupoles which we will further discuss in sec. \ref{sec: GWs}. 
In comparison with the  analytic result  (\ref{ri}), 
the simulations show that departures from the analytic description start appearing for times  $t>t_\text{max}$. These departures might be mild or strong depending on the deformation parameters $(\alpha, \beta, \gamma)$.  $N$-body simulations are  particularly valuable for times $t>t_\text{col}$, where the analytical solutions (\ref{ri}) are definitely not applicable.

\begin{figure}[!htbp]
  \begin{subfigure}{.5\textwidth}
  \centering
  \includegraphics[width=1.0 \linewidth]{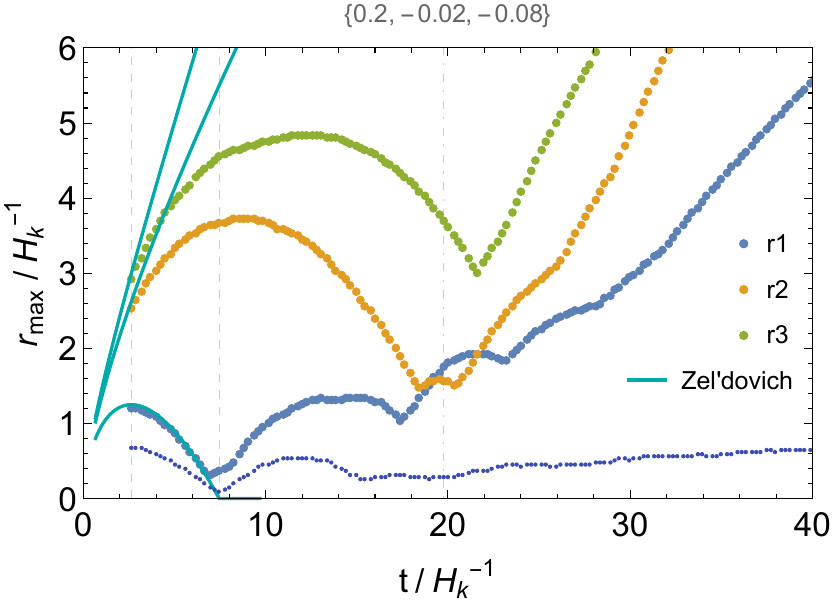}
\end{subfigure}  \quad\quad
  \begin{subfigure}{.5\textwidth}
  \centering
  \includegraphics[width=1\linewidth]{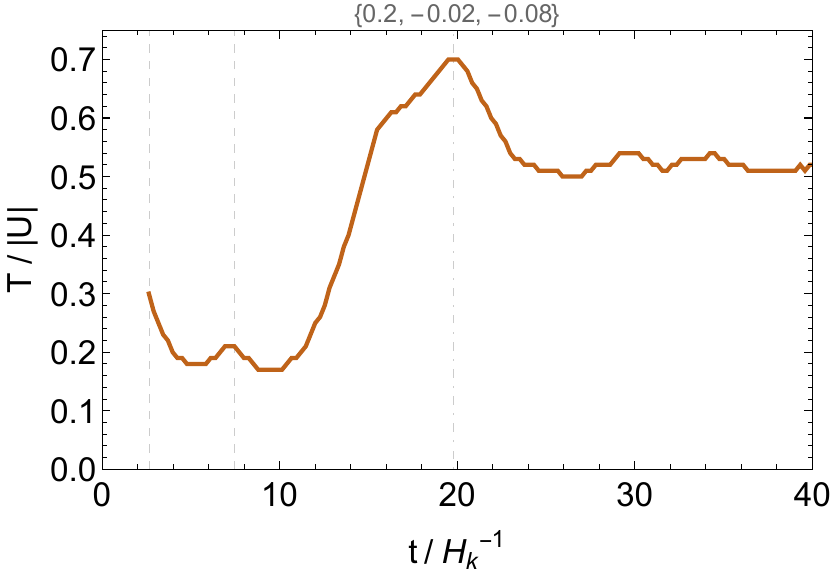}
\end{subfigure} \\
\begin{subfigure}{.5\textwidth}
  \centering
  \includegraphics[width=1.0 \linewidth]{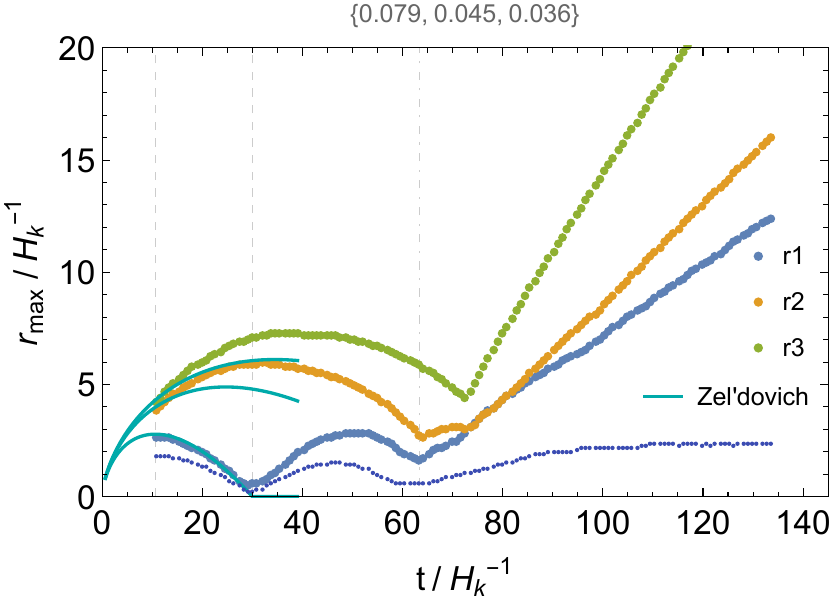}
\end{subfigure}  \quad\quad
  \begin{subfigure}{.5\textwidth}
  \centering
  \includegraphics[width=1\linewidth]{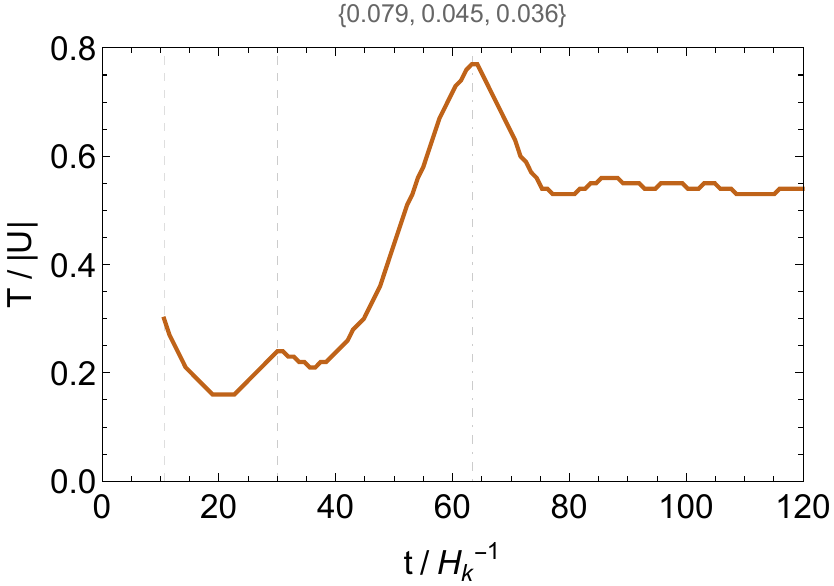}
\end{subfigure}
 \begin{subfigure}{.5\textwidth}
  \centering
  \includegraphics[width=1.0 \linewidth]{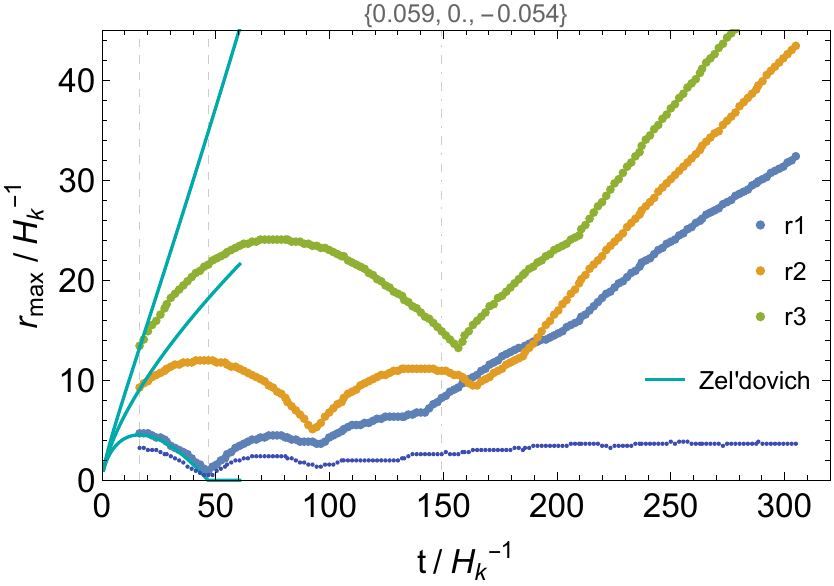}
\end{subfigure}  \quad\quad
  \begin{subfigure}{.5\textwidth}
  \centering
  \includegraphics[width=1\linewidth]{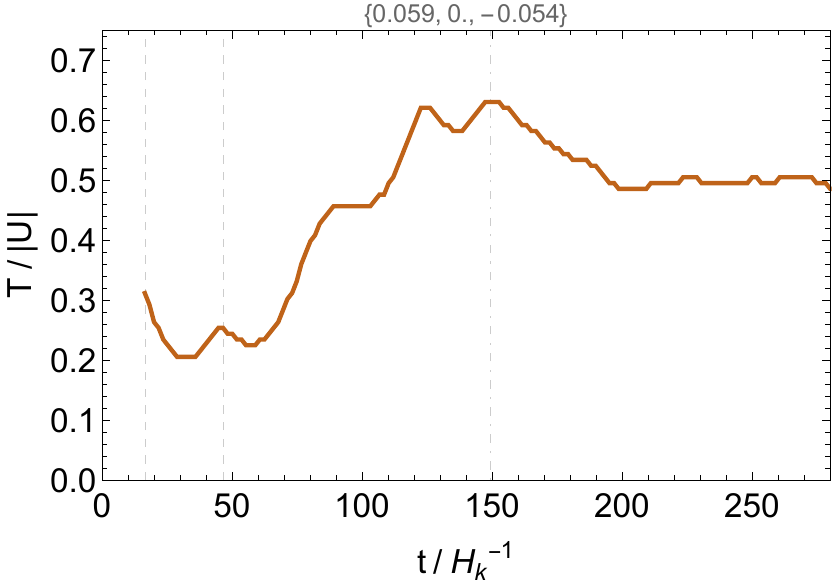}
\end{subfigure}

 \caption{\label{fig: Vir}~~  
$N$-body results are displayed for an overdensity with $\sigma=0.1$ and deformation parameters 
$(\alpha, \beta, \gamma)=(0.2, -0.02, -0.08)$ ({\it top panels}), 
$(0.079, 0.045, 0.036)$ ({\it middle panels}), $(0.059, 0, -0.054)$  ({\it bottom panels})
  simulated by $N=2500$ particles.
{\it Left  panels}:  The coordinate of the outermost particles in the $N$-body distribution along the directions $r_1$, $r_2$ and $r_3$ are shown. With the thin dotted blue line the evolution of an inner shell 
is shown. For comparison the Zel'dovich solution is depicted in cyan lines.
{\it Right panels}: The virial ratio of kinetic over potential energy of the $N$-body distribution.  
The gridlines in all the plots denote  from left to right the moments of turnaround ($t_{\rm max}$), pancake collapse ($t_{\rm col}$) and the bottleneck stage ($t_{\rm virial}$).
}
\end{figure}

A system of $N$ particles interacting gravitationally with total mass $M$ and a characteristic
dimension $R$ 
reaches a dynamic equilibrium state
on a timescale comparable to a few times the typical free fall timescale 
$t_\text{ff}\, \approx \,1/\sqrt{GM/R^3} \sim (G\rho)^{-1/2}$, where $\rho$ is the mass density of the distribution.
This is the response time needed to settle down to virial equilibrium, that is
\begin{align} \label{vir}
\frac{T}{|U|} = 0.5\,,
\end{align}
where $ T=\frac{1}{2} \sum_{i=1}^N m_i  |\vec{v}_i|^2$  
is the kinetic  and
$U=-\frac12 \sum_{i\neq j} Gm_i m_j/ |\vec{x}_i- \vec{x}_j|$ the potential energy of the system.
Equation (\ref{vir}) is a statement of the scalar virial theorem.
If the system is initially out of equilibrium, virialization is reached through mixing in phase space due to fluctuations of the gravitational potential, a process called violent relaxation 
\cite{LyndenBell:1966bi}.
A reasonable approximation is that a  spherical distribution virializes at about the same time that this idealized
model collapses to a singular density. For  anisotropic collapse the simulations show that the virialization time takes longer,  about a couple of $t_\text{col}$ times, see the right plots of fig. \ref{fig: Vir}. 
 After the distribution has settled to approximate equilibrium, its radius can be
estimated from the virial theorem. 
An analytical result exists for a spherical configuration 
with the half-mass radius of the distribution being about one-third of the turnaround radius. 
For anisotropic collapse, the results of the simulation show that roughly the same conclusion holds.

The initial conditions for our $N$-body simulations are critical and we  determine them by the Zel'dovich solution. 
 We assume that the Zel'dovich approximation describes sufficiently well the evolution of the overdensity until the moment of maximum expansion $t_\text{max}$. 
After that  moment  we assume that the overdensity has effectively decoupled from the background and the Hubble flow can be neglected from the gravitational dynamics. 
We  consider the  evolution of the overdensity for times $t\geq t_\text{max}$ as an effectively cold gravitational collapse and we describe it utilizing the $N$-body simulations.

\begin{figure}[t!]
  \begin{subfigure}{.323\textwidth}
  \centering
  \includegraphics[width=1 \linewidth]{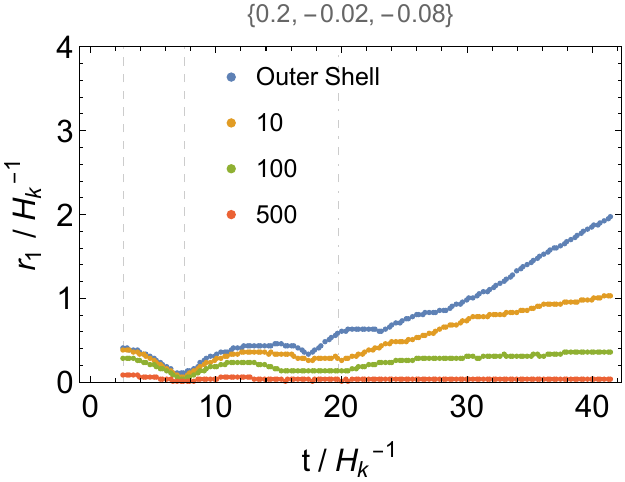}
\end{subfigure} 
  \begin{subfigure}{.323\textwidth}
  \centering
  \includegraphics[width=1\linewidth]{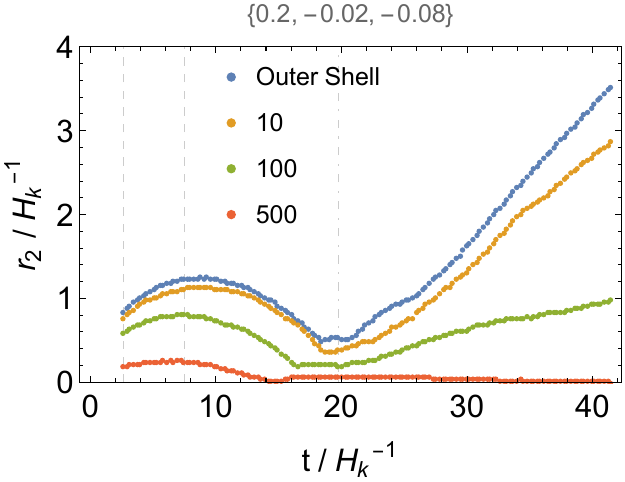}
\end{subfigure}
\begin{subfigure}{.323\textwidth}
  \centering
  \includegraphics[width=1\linewidth]{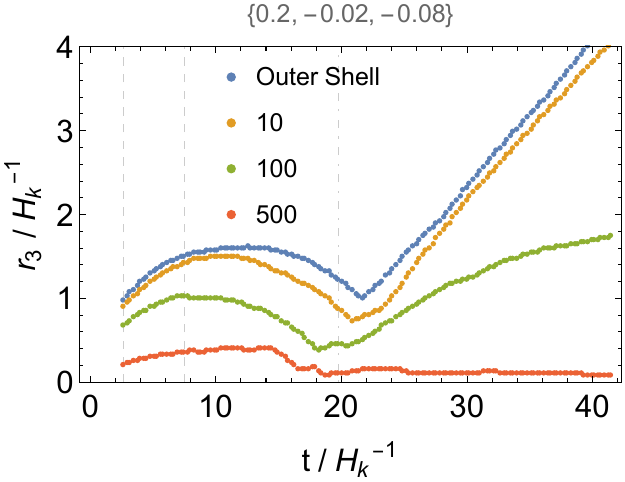}
\end{subfigure}
 \caption{\label{fig: shells}~~  
The panels  show the evolution of the semi-axes of the outer and three inner one-particle shells (10th, 100nd, 500th) of the $N$-body distribution in directions $r_1, r_2$ and $r_3$ for  $\sigma=0.1$ and deformation parameters 
$(\alpha, \beta, \gamma)=(0.2, -0.02, -0.08)$.
  }
\end{figure}

We suppose that at some initial time $t_k$
  there is a density fluctuation of size $k^{-1}$ and variance $\sigma^2=\left\langle \delta^2_{\rm L} \right\rangle$, with a small initial asphericity described by nonzero values for  ($\alpha, \beta, \gamma)$, that  enters the Hubble radius.
In the framework of $N$-body simulation the density perturbation is 
initially composed of  an approximately  homogeneous distribution of identical  $N$ particles. In analogy with eq. (\ref{ri}) 
the semi-axis of the ellipsoidal particle distribution $r_{i}^{\rm NB}(t)$ can be numerically specified.
We run the  $N$-body simulation for initial conditions that are set at the moment of the turnaround.
Hence, the initial positions $|\vec{r}^{\,\rm NB}_{\rm init}|$ of the particles placed at the boundary of the
  distribution and  along the directions of the principal axes 
   are determined by eq. (\ref{ri}) and (\ref{norm}) for $t= t_\text{max}$, 
\begin{equation} \label{rinit}
r_{1\,, \rm init}^{\rm NB} =\frac{1}{4\,\alpha } H_k^{-1}
\,, \quad \quad
\frac{r_{2 \,, \rm init}^{\rm NB}}{r_{1 \,, \rm init}^{\rm NB}}=2- \frac{\beta}{\alpha} \,, \quad \quad
\frac{r_{3 \,, \rm init}^{\rm NB}}{r_{1 \,, \rm init}^{\rm NB}}=2- \frac{\gamma}{\alpha}\,.
\end{equation}
Accordingly, the maximal initial velocities $v_i$ along the directions of the principal axes are
\begin{equation} \label{vinit}
v_{1 \,, \rm init}^{\rm NB}=0, \quad\quad
v_{2 \,, \rm init}^{\rm NB}= \sqrt{2}\,\,\frac{\alpha-\beta}{\sqrt{\alpha}}\, , \quad\quad
v_{3 \,, \rm init}^{\rm NB}=\sqrt{2}\,\,\frac{\alpha-\gamma}{\sqrt{\alpha}}\,.
\end{equation}

We study systems with  $N \lesssim 10^4$ particles each with equal mass $m_i=M/N$. 
All systems are initialized by randomly placing particles in an ellipsoid volume with  space borders prescribed by the 
equation $(x/r_{1 \,, \rm init}^{\rm NB})^2+ (y/r_{2 \,, \rm init}^{\rm NB})^2+(z/r_{3 \,, \rm init}^{\rm NB})^2=1$
and radial velocities that 
increase linearly from zero, in the center of the distribution, to the 
boundary values given by eq. (\ref{vinit}). 
With these initial conditions, the  distribution evolves anisotropically under the influence of the self-gravitational field.
In our numerics we set $G=1$. 
 We calibrate
  the timescale of the $N$-body evolution by the timescale $t_\text{max}$ of Zel'dovich dynamics
  by fixing 
  the span of $N$-body evolution from the initial time  until the pancake collapse as follows
\begin{equation}
\Delta t_\text{col}=(t_\text{col}-t_\text{max}) \approx 1.8 \, t_\text{max}\,.
\end{equation}  
This is approximately  equal to the free-fall time $\sim (G\rho)^{-1/2}$ for a particle at the position $r_{1 \,,\rm init}^{\rm NB}=(3t_\text{max}/2)^{2/3}/2$.
The mass is assumed to be constant and cold throughout this first stage of collapse, i.e. 
there is no flow of matter across the shells. 
The fact that we follow the Zel'dovich approximation until the moment of maximum expansion might introduce a systematic error in our results, which presumably it is a minor one.  
 The advantage of  this methodology  is that it makes our analysis  self-consistent and transparent   
and any refinement to the overall signal can be 
consistently carried out.

We numerically track the evolution of the system   from the moment of turnaround until times $t_\text{end}\gg t_\text{max}$ so that the collapse, the violent relaxation and virialization processes take place.
Within the time interval $[t_\text{max}, t_\text{end}]$  we
take successive snapshots of the  $N$-body distribution with step $\Delta t_{\rm step}^{\rm NB} \ll t_\text{max}$.
We stop the $N$-body simulation deep in the virialized regime, where a nearly spherical configuration is formed surrounded by an outflow of  particles escaping the gravitational potential, see figs. \ref{fanim} and \ref{fig: shells}.
We take a common $t_\text{end}$ value for every configuration  $t_{\rm end} \sim 150\, t_k$.
Roughly   500 snapshots from $t_\text{max}$ until  the end time 
$t_{\rm end}$
are sufficient to capture the details of the evolution of the $N$-body distribution. 
Those configuration 
that evolve slowly and virialize later than $150\, t_k$,   give a negligible contribution to the  GW signal.
Deviations from Zel'dovich description are observed already  at the stage of collapse with the 
radii of the $N$-body distributions having much smaller sizes compared to those given by eq. (\ref{ri}),
\begin{equation}
r^{\rm NB}_2(t_\text{col}) < r^{\rm Zel}_2(t_\text{col})
 \,, \quad 
r^{\rm NB}_3(t_\text{col}) < r^{\rm Zel}_3(t_\text{col}) \,. 
\end{equation}
This means that the bulk mass is much more concentrated compared to the Zel'dovich picture at the time of pancake collapse. The deviations decrease for  the case where $\gamma\sim\beta\sim\alpha$ where a nearly spherical collapse takes place.
Eventually,  the Zel'dovich approximation breaks down and 
for $t\gtrsim t_\text{col}$ the $N$-body simulations reveal the subtle details of the 
 gravitational evolution.
 
For illustration reasons
we choose three benchmark  configurations, out of several thousands examined,  to explicitly display the evolution of the collapsing distribution, 
 see figs. \ref{fig: Vir}.
In these benchmark configurations some of the subtle features in the quadrupole evolution are clearly evident and the deviations from Zel'dovich approximation are manifest. 
 In  the  left panels of  figs. \ref{fig: Vir}
 the  evolution of the $r_i^{\rm NB}$ is tracked by the  particles positioned at the boundary of the distribution along $r_1$, $r_2$ and $r_3$ directions respectively.  
 The result of the $N$-body simulation, depicted with dots separated with time step $\Delta t_{\rm step}^{\rm NB}$, 
 is compared  
with the Zel'dovich solution (solid cyan lines). 
The moments of  turnaround and pancake collapse are  indicated with dashed gridlines, while  the so-called bottleneck stage is indicated with the dot-dashed gridline.
 At $t_{\rm col}$ and along $r_1$ direction the Zel'dovich solution  is in agreement with  the $N$-body result by construction but in the other two directions there is an apparent discrepancy. In the plot we also see that for $t\gg t_{\rm col}$  the outermost particles of the distribution finally escape the central concentration. Clustering occurs mainly for inner particles. In the plots the thin dotted blue line, that corresponds to 200th outermost particle along $r_1$, illustrates the evolution of an inner one-particle shell. 
 The evolution of inner shells is displayed in more detail in fig. \ref{fig: shells} where the outflow and inner flow motions are demonstrated.

Our objective is  the  computation of 
the average GW energy emitted by 
a Hubble patch of size $k^{-1}$ that undergoes gravitational collapse and 
the finding of the total  spectrum of GWs propagating 
in our Universe today.
Taking a particular combination of deformation parameters $(\alpha, \beta, \gamma)$ is only a case study out of infinite 
combinations 
and does not, of course, represents the average  GW emission. Actually, the 
differences in the emissivity   of GWs can be extremely large for different values of the deformation parameters. 
This means that we have to consider a large number of different configurations  $N_\text{conf}$, each characterized by a unique combination $(\alpha, \beta, \gamma)$, 
and properly add their gravitational contribution. 
This is done by considering the Doroshkevich PDF ${\cal F}_{\rm D}(\alpha, \beta, \gamma, \sigma)$ which gives
the statistical significance of a perturbation with a specific asphericity.

\subsection{ Doroshkevich PDF} \label{subsec: Dor}

The average GW signal is the cumulative effect of numerous gravitational sources and  has a critical dependence on the statistical properties of the deformation parameters.
We assume that the shape-fluctuation defines a three-dimensional random field that follows a Gaussian distribution. 
For a homogeneous and isotropic background  space   
 the probability distribution function  for $\alpha$, $\beta$ and $\gamma$ is
\begin{align}
{\cal F}_\text{D}(\alpha, \beta, \gamma) d\alpha d\beta d\gamma =& -\frac{27}{8\sqrt{5}\pi \sigma^6_3}\times \nonumber 
 \exp\left[- \frac{3}{5 \sigma_3^2}\left((\alpha^2+\beta^2 +\gamma^2)-\frac{1}{2}(\alpha\beta+\beta\gamma+\gamma\alpha) \right)\right] \nonumber \\
& \times (\alpha-\beta)(\beta-\gamma)(\gamma-\alpha) d\alpha d\beta d\gamma,
\end{align}
and  is called  Doroshkevich PDF \cite{Doroskevich1970}. Doroshkevich was the first to apply these methods extensively to the study of the formation of cosmic structure.
The $\sigma^2_3$ is the dispersion of the diagonal components 
of $D_{ij}$.
The value $\sigma_3^2$ is related to the dispersion of the density perturbations by 
\begin{align} \label{vevdelta}
 \sigma^2  \equiv \left \langle \delta^2_{\rm L} \right\rangle & =  \,  \int_{-\infty}^{\infty}\,d\alpha \int_{-\infty}^{\alpha}d\beta \int_{-\infty}^{\beta} d\gamma
\, 
 {\cal F}_\text{D} (\alpha, \beta,\gamma, \sigma_3)
\left({\alpha+\beta+\gamma}\right)^{2} \,= \,5 \,\sigma_3^2\,.
\end{align}

At the early time $t_k$ the density perturbation in each Hubble patch is characterized by a degree of initial asphericity
described by the parameters $(\alpha, \beta, \gamma)$ of the deformation tensor with
distribution of the values  given by ${\cal F}_\text{D}(\alpha, \beta, \gamma, \sigma)$ probability density function (PDF).  
The probability two principal values of the deformation tensor to be nearly equal ($\alpha \simeq \beta \simeq \gamma$) is particularly small. 
Extreme values of asphericity are also rare.
Along the lines of eq. (\ref{vevdelta}) the quantities $\left\langle t_{\rm max}\right\rangle$ and $\left\langle t_{\rm col}\right\rangle$ can also be found.

The differential number of sources with deformation parameters lying in the range $(\alpha, \alpha+d\alpha), (\beta, \beta+d \beta)$ and $(\gamma,\gamma+d\gamma)$ are
\begin{align}
dN_\text{source}= \frac{ {\cal H}^{-3}_0}{ k^{-3}} {{\cal F}_\text{D}(\alpha, \beta, \gamma, \sigma)\, d\alpha d\beta d\gamma}\,,
\end{align}
where ${\cal H}_0$ is the comoving Hubble parameter today.
The Doroshkevich PDF is normalized to unity and   integration over the entire parameter space $(\alpha, \beta, \gamma)$ gives $N_\text{source}$ equal to ${\cal H}^{-3}_0/ k^{-3}$, which  is the number of Hubble patches of size $\frac{4}{3}\pi k^{-3}$ enclosed in the  comoving volume of the present Universe. 
Here we are interested in patches that enclose an overdensity hence we restrict ourselves in integration intervals  dictated  by eqs. (\ref{hiera}) 
and, additionally, by the requirement $t_\text{max} \geq t_k$. 
The last condition reads  $\alpha \leq 1/2$ and together with eqs. (\ref{hiera}) they define the parameter space of interest  ${\cal S}$. 
The integral $I_{\rm D}= \int\int\int_{\cal S}
\,d\alpha d\beta d\gamma\, {{\cal F}_\text{D} (\alpha, \beta,\gamma, \sigma)}$ is to a good
approximation equal to $1/2$
Besides the number of sources 
 ${ {\cal H}^{-3}_0}{ k^{3}} I_{\rm D}$ 
 we will also consider an ensemble  of configurations $N_{\rm conf}$ in order to estimate the average GW signal emitted by a single Hubble patch.  The size of $N_{\rm conf}$ is  determined by the number of bins with width $\Delta \alpha$, $\Delta \beta$ and $\Delta \gamma$ that span the parameter space ${\cal S}$ of possible deformations. 
We will discuss the significance of $N_{\rm conf}$ and the specification of the optimal number of bins  in more detail  in sec. \ref{sec: average}.


\section{The energy density of GWs emitted during gravitational collapse } \label{sec: GWs}

As the overdensity undergoes rapid compression and collapses under its gravity, the asymmetrical process generates gravitational waves.
 By studying the properties of the emitted waves we can learn about the presence of possible overdensities,  their masses and amplitudes, their nonlinear evolution and probe the  history of the early Universe.
We estimate the spectrum of GWs emitted from mass distributions that end up  to halo formation via $N$-body simulations. We also compare it with analytic results, that partially describe the collapsing process.
We use  
 the quadrupole approximation and  assume that the quadrupole moment is the primary source of gravitational radiation, while higher-order moments 
 have significantly weaker contributions.

\subsection{Quadrupole approximation}

The equation of motion of tensor perturbations to a background of a Friedmann-Robertson-Walker metric in conformal Newtonian gauge is
\begin{equation}
\ddot{h}_{ij} +3H  \dot{h}_{ij} - \frac{\nabla^2}{a^2} h_{ij}=4 S_{ij}^{\rm TT} \,,
\end{equation}
where  $S_{ij}^{\rm TT}$ is the source term, the transverse-traceless (TT) part of the anisotropic stress tensor \cite{Ananda:2006af, Baumann:2007zm, Malik:2008im}.
The total gravitational energy radiated is distributed among multipoles  of the perturbation $h_{\mu\nu}$. 
To lowest order in the multipole expansion it is $h_{ij}(t, d) \propto \ddot{Q}_{ij}(t-d/c)/d$, where $Q_{ij}$ is the mass quadrupole moment, $d$ the distance from the source and the overdot denotes time derivative.
This formula is valid for any nearly Newtonian slow-motion source. 
 Such sources  emit GWs with wavelength larger than  the size of the source, that implies  internal velocities  $v\ll 1$ \cite{Sathyaprakash:2009xs}. 
In terms of the moment of inertia the  quadrupole moment is written as
\begin{equation} 
Q_{ij}=-I_{ij}(t)+ \frac13 \delta_{ij} \text{Tr} I(t),
\label{Qij}
\end{equation}
where $I_{ij}$, is the moment of inertia tensor  defined as $I_{ij}=\int d^3r \, \rho ( \vec{r}) \left(\delta_{ij}|\vec{r}|^2- r_i r_j  \right)$.
The anisotropic collapse of the bulk mass of the density perturbation
produces gravitational radiation.
The power emitted in the form of GWs from a Hubble patch, that encloses a perturbation with a quadrupole tensor $Q_{ij}$, is
\begin{equation} \label{Pdt}
\frac{dE_{\rm GW}}{dt}=\frac{G}{5c^5} \sum_{ij} \dddot{Q}_{ij}(t)\dddot{Q}_{ji}(t).
\end{equation}
After expanding the quadrupole in Fourier series and  taking the continuum limit for $\omega$, where $\tilde{Q}_{ij}$ is the Fourier mode of the quadrupole $Q(t)$, the spectral density of GWs 
 is found to be
\begin{align}  \label{result_Pd}
& \frac{dE_{\rm GW}}{d\ln \omega}=\frac{4\pi G}{5c^5} \omega^7 \sum_{ij}|\tilde{Q}_{ij}(\omega)|^2.
\end{align}
Gravitational radiation  emitted at time $t_{e}$ with an energy $E_{\rm GW}$ is today observed with energy  $E_{\rm GW}/(1 + z_{\rm e})$ and angular frequency $\omega_0 = \omega/(1 + z_{\rm e})$. The cosmological redshift
  is given by $1 + z_{\rm e} = a_0/a(t_{\rm e})$, where we take $a_0 = 1$.
The differential energy density parameter of the stochastic GW background per observed logarithmic frequency
${\rho_\text{crit}}^{-1} {d\rho_\text{GW}}/{d \ln \omega}$ is given in ref.
\cite{Dalianis:2020gup}
\begin{align}
\Omega_\text{GW}(t_0,\omega_0) =\frac{1}{\rho_\text{crit}(t_0)}
  \int\int\int_{\cal S}
\,d\alpha d\beta d\gamma\, \frac{1}{1+z_{\rm e}} \frac{1}{V_k}\frac{4 \pi G}{5 c^5}  \sum_{ij}|\tilde{Q}_{ij}\left(\omega \right)|^2 \omega^7
 {{\cal F}_\text{D} (\alpha, \beta,\gamma, \sigma)}\,,   \label{result_Om}
\end{align} 
where the integration  takes place over the parameters space of interest  ${\cal S}$. 
$\rho_\text{crit}(t_0)$ is the critical energy density of the Universe today and $V_k=4\pi k^{-3} /3$ is the comoving volume  at the time of horizon entry of the perturbation.
 In  terms of the frequency parameter today $f_0$,
Eq. (\ref{result_Om}) is recast
via the relation $\omega=2\pi f_0(1+z_{\rm e})$.
We will assume that the EMD era 
 transitioned abruptly into RD era at the cosmic time $t_{\rm rh}$, where the subscript ``${\rm rh}$" denotes quantities defined at the moment of reheating. 
 In this case  
the comoving wavenumber for $k>k_{\rm rh}$  and the comological redshift are 
\begin{align}\label{kMgen}
  k(M,  T_\text{rh}) \simeq \left(\frac{3M(1+z_\text{rh})^3}{4\pi \rho_\text{rh}} \right)^{-1/3}\,,
\quad
\quad
  1+z_{\rm e}=\frac{1+z_\text{rh}}{(6\pi G \rho_\text{rh})^{1/3}} \, t_{\rm e}^{-2/3} \,,
 \end{align}
where  $\rho_\text{rh}=\pi^2 g_* T_\text{rh}^4/30$ and $z_{\text{rh}}$  are respectively the energy density and the redshift at the time of reheating.
Otherwise, in the case of a gradual transition from the EMD to RD, there will be deviations from the approximation $w=0$ and the expansion rate will be different depending on
the fractional distribution of the energy density components  (as e.g. in the approach of \cite{Fernandez:2023ddy}).   Here we will focus on the ideal case of cold gravitational collapse 
which is clear  both in the modeling and the simulation level.

The expression (\ref{result_Om}) is the equation which gives the total average signal produced in a $\Delta t=t-t_k$ time interval during which   
$\dddot{Q}_{ij}(t)\neq 0$.  
Finding  $Q_{ij}$ is the ultimate objective and  we will determine utilizing $N$-body simulations. For the sake of comparison
we will first review   the result found in the framework of  Zel'dovich approximation \cite{Dalianis:2020gup} in the 
following subsection.

\subsection{GWs computed from the Zel'dovich solution} \label{GWzel}
Within  Zel'dovich approximation we get the solutions (\ref{ri}) and 
the evolution of  $Q_{ij}(t)$ can be described analytically for times $t_k \lesssim t\lesssim t_{\rm col}$. 
 By choosing  the principal axes frame, the moment of inertia tensor of the ellipsoid is $I_{ij}=\frac15 M {\rm diag}(r^2_2+r_3^2, r_1^2+r_3^2, r^2_1+r_2^2) $,
where  $M$ is the total enclosed mass, and the components of the quadrupole are 
\begin{align}
    Q^{\rm Zel}_{ij}(t)=\frac{3M}{80} \left(\frac{t_k}{t^2_\text{max}}\right)^{2/3} \frac{1}{\alpha^2}
    \left[\left(3c^2_{ij} -\alpha^2-\beta^2-\gamma^2\right)t^{8/3} \right. 
     \left. -4 \left(3c_{ij} -\alpha-\beta-\gamma \right)t^2 \, t^{2/3}_\text{max} \right] \delta_{ij}\,,
\end{align}
where $c_{ij}={\rm diag}(\alpha, \beta, \gamma )$. 
 In particular for a time bin $[t_1, t_2]$ within the $[t_k, t_\text{col}]$ interval  the analytic expression is found
\begin{align} \nonumber
   \omega^6 \sum_{ij}|\tilde{Q}^{\rm Zel}_{ij}  (\omega)|^2 = &\frac{1}{54 \pi^2} \frac{t_k^{4/3}t_1^{4/3}}{t_\text{max}^{8/3}}M^2
     \left[1+\left(\frac{\beta}{\alpha}\right)^4+\left(\frac{\gamma}{\alpha}\right)^4
    -\left(\frac{\gamma}{\alpha}\right)^2-\left(\frac{\beta}{\alpha}\right)^2\left(1+\left(\frac{\gamma}{\alpha}\right)^2\right)\right]  \nonumber \\
    & \times \left|\text{Ei}\left[\frac13, i\omega t_1\right]- \left (\frac{t_2}{t_1} \right )^{2/3} \text{Ei}\left[\frac13, i\omega t_2\right] \right|^2 \,.
    \label{FT2}
\end{align}
The above expression makes possible the exact calculation of the integral (\ref{result_Om}). 
In Ref. \cite{Dalianis:2020gup}  the GW signal 
is computed for a range of frequencies after a straightforward numerical  integration, under the approximation (\ref{ri}) valid for times $ t \lesssim  t_\text{col}$.  
The only required  input  is  the deviation of the density perturbations at horizon entry in the linear regime, $\sigma$, and the characteristic length scale of the density perturbation $k^{-1}$ (or the mass $M$) which specifies the time of entry  $t_k$.

\begin{figure} 
  \begin{subfigure}{.48\textwidth}
  \centering
  \includegraphics[width=1.0 \linewidth]{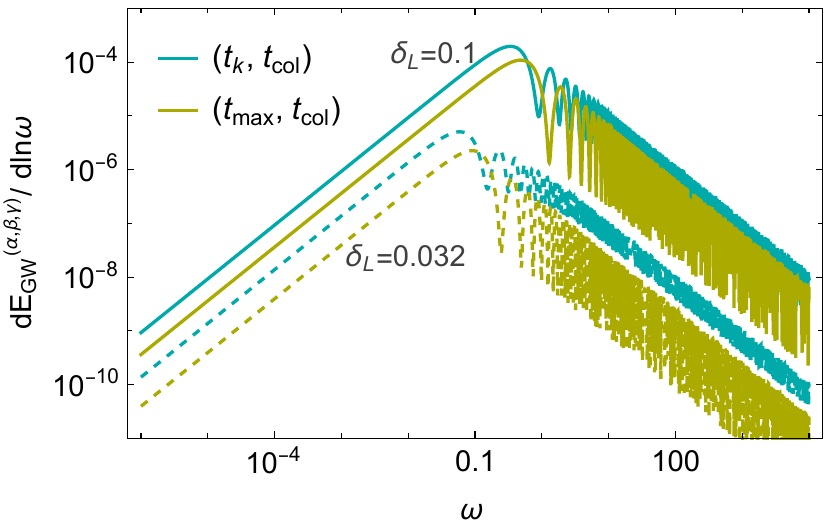}
\end{subfigure}  \quad\quad
  \begin{subfigure}{.48\textwidth}
  \centering
  \includegraphics[width=1\linewidth]{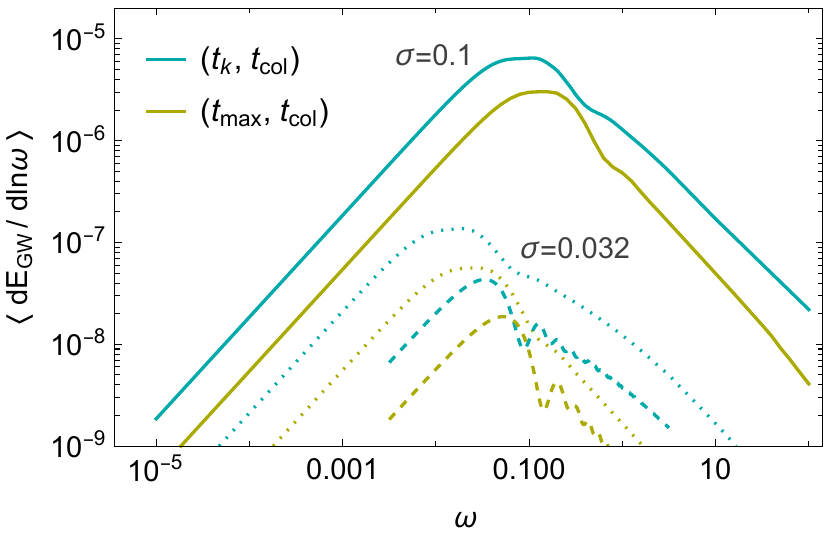}
\end{subfigure}
 \caption{\label{Zeld1}~~  
{\it Left  panel}: The Zel'dovich solution for the GW spectrum for a specific example of deformation parameters
$(\alpha, \beta, \gamma)=(0.2, -0.02, -0.08)$  (solid lines) and 
$(0.2/\sqrt{10}, -0.02/\sqrt{10}, -0.08/\sqrt{10})$  (dashed lines) which correspond to $\delta_{L}=0.1$ and $\delta_{L}=0.032$ respectively. 
 The dark cyan lines are for the interval $(t_1, t_2)=(t_k, t_{\rm col})$ and dark yellow for $(t_1, t_2)=(t_{\rm max}, t_{\rm col})$.
{\it Right Panel}: The average GW spectrum found after integration over the  parameter space ($\alpha, \beta, \gamma$) with the constraint $t_{\rm max} < 50\, t_k$ for  $\sigma=0.1$ (solid line) and $\sigma =0.1/\sqrt{10}\approx0.032$ (dashed line). 
The dotted lines correspond to an integration with the constraint $\left\langle t_{\rm max} \right\rangle <50 t_k$.
In the plots we have set $H_k=1$.
}
\end{figure}

The spectrum of the emitted energy $E_{\rm GW}$ is given by  eq. (\ref{result_Pd}) and involves exponential functions ${\rm Ei}$ with an argument that depends on time, either the initial  $t_1$ or the final time $t_2$,  described by eq. (\ref{FT2}). One choice is to consider the time interval $(t_1, t_2)=(t_{\rm max},t_{\rm col})$. Another choice it is to take the limit $t_1\rightarrow t_k$ that gives the quadrupole evolution 
 from the moment of horizon entry. 
These two choices yield spectra with small differences as one can see in the left panel in fig. \ref{Zeld1}. 
At IR frequencies $\omega t_2 \lesssim 1$
the terms ${\rm Ei}[1/3, i\omega t_1]$ and $(t_2/ t_1)^{2/3}{\rm Ei}[1/3, i\omega t_2]$ balance each other and the expression (\ref{FT2}) is
constant.
These  imply that the quantity  $ \omega^7 \sum_{ij}|\tilde{Q}^{\rm Zel}_{ij}  (\omega)|^2$ 
has a peak at $\omega \sim t_{2}^{-1}$ with amplitude proportional to $\delta_{\rm L}^{7/2} /t_k$.
In the left panel of fig. \ref{Zeld1}  four GW spectra are depicted for $H_k=1$.
With dark yellow lines the results of ref. \cite{Dalianis:2020gup} are depicted, whereas the dark cyan lines correspond to the choice $(t_1, t_2)=(t_{k}, t_{\rm col})$ that we consider in the current analysis.

\begin{figure} 
  \begin{subfigure}{.48\textwidth}
  \centering
 \includegraphics[width=1\linewidth]{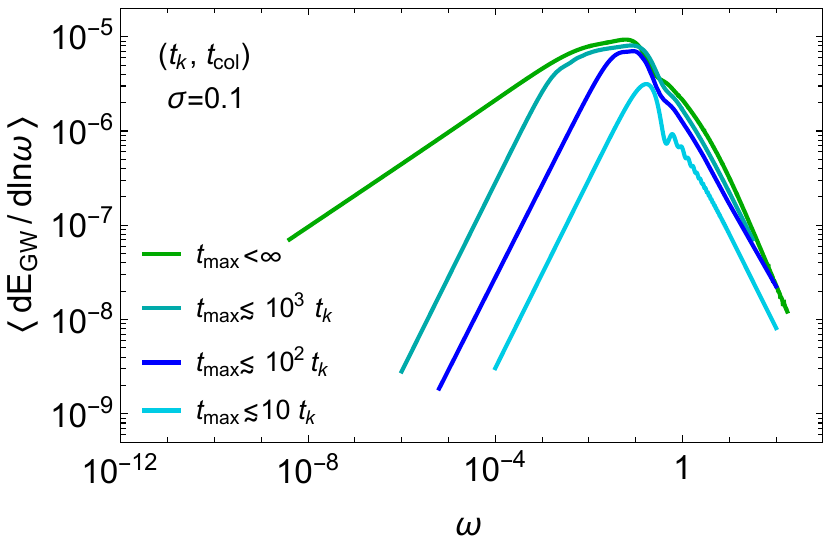}
\end{subfigure}  \quad\quad
  \begin{subfigure}{.48\textwidth}
  \centering
 \includegraphics[width=1\linewidth]{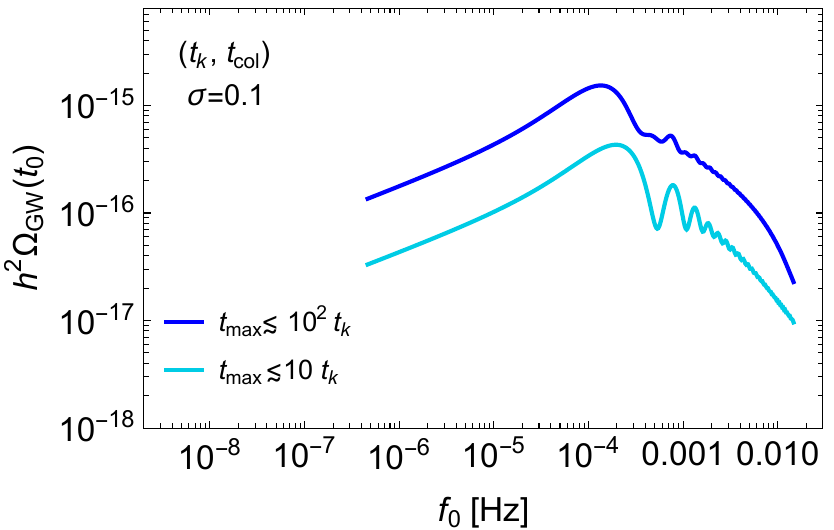}
\end{subfigure}
 \caption{\label{ZeldintIR}~~  
{\it Left panel:} The average GW power emitted from configurations that experience a turnaround at  $t_{\rm max} \lesssim 10 t_k$, $t_{\rm max} \lesssim 10^2 t_k$,  $t_{\rm max} \lesssim 10^3 t_k$ respectively from the bottom to the top. The upper curve is the power emitted for arbitrary large times including  configurations that  turnaround at cosmologically non-viable times. {\it Right Panel}: The corresponding differential energy density parameter in the present Universe $\Omega_{\rm GW}(t_0, f_0)$  for $M=10^{-12}M_\odot$ and $T_{\rm rh}=1$ GeV.}
\end{figure}

The average GW energy emitted  is given, within the Zel'dovich approximation, by the integral 
\begin{align}  \label{vevPd}
\left\langle \frac{dE^{\rm Zel}_{\rm GW}}{d\ln \omega}\right\rangle & =\int\int\int_{\cal S}  
\,d\alpha d\beta d\gamma\,  
  \frac{4 \pi G}{5 c^5}  \omega^7 |\tilde{Q}^{\rm Zel}_{ij}\left(\omega)\right)|^2 
 {{\cal F}_\text{D} (\alpha, \beta,\gamma, \sigma_3)} \,.
 \end{align}
The exact value of the peak depends on the interval of integration $\alpha_{\rm min}<\alpha<1/2$. 
By taking smaller $\alpha_{\rm min}$ values the amplitude of the average GW spectrum receives contributions from configurations with slow evolution and the amplitude of the spectrum increases, particularly in the infrared frequency range, see fig. {\ref{ZeldintIR}}. For the limiting case that $\alpha_{\rm min} \rightarrow 0$  configuration that experience a turnaround at infinite times are also counted. 
  Because of the finite duration of the EMD phase we consider $\alpha_{\rm min}$ values which give $t_{\rm max} < 10^2 t_k$ or $\alpha > 0.023$.
For larger values for $\alpha_{\rm min}$ an oscillatory pattern appears in the  ultraviolet regime of the spectrum, see the bottom curve in fig.  {\ref{ZeldintIR}}.  This can be understood by the fact that larger $\alpha_{\rm min}$ limits the spread of $t_{\rm max}$ values and consequently  it operates as a time-filter selecting configurations which evolve in more synchronized manner. 
Equivalently,  the oscillatory pattern also appears by keeping $\alpha_{\rm min}$ constant but decreasing the $\sigma$ value.
For $t_{\rm max} < 10^2 t_k$    we attain the plots depicted in the right panel of fig. \ref{Zeld1}. 
   The spectrum peaks at the angular frequency $\omega_{\rm peak} \sim t_{\rm col}^{-1}$ 
   with maximum amplitude   $ \sim 8 \times 10^{-6}$ for $\sigma=0.1$ and having set $H_k=1$. 
   By decreasing $\sigma$ the amplitude of the spectrum decreases in a  way that depends on the parameter space of integration ${\cal S}$. If the ratio $t_{\rm max}/t_k$ is kept fixed  while $\sigma$ gets smaller the ensemble of configurations with a significant PDF value shrinks, which  changes both the amplitude and the shape of the spectrum (depicted with dashed lines in the r.h.s. of fig. \ref{Zeld1}).
If  instead  we decrease $\sigma$ but keep  the $\sigma$-dependent  ratio $\left\langle t_{\rm max} \right\rangle/t_k$ fixed, 
 the shape of the spectrum nearly does not alter: it  decreases roughly as $\sigma^{7/2}$  with the peak shifting at smaller frequencies according to the relation $\omega_{\rm peak}\propto \sigma^{3/2}$ (dotted lines in the r.h.s. of fig. \ref{Zeld1}).

The expression (\ref{vevPd}) gives the average GW power at the time of emission  from a perturbation of an arbitrary mass.   In order to find  the signal received in the present Universe a mass scale for the overdensity and $T_{\rm rh}$ have to specified.
 The  differential energy density parameter of the stochastic GW background $\Omega_{\rm GW}$ can be computed after choosing values for the reheating temperature $T_{\rm rh}$ and 
  the mass $M$ of the perturbation:
\begin{align}\label{result_OmZ}
\Omega^{\text{Zel}}_\text{GW}(t_0,f_0) & \, = \,
\frac{2^4}{30 \times 54}\,
\frac{g_s (T_0)\, T_0^3}{\rho_\text{crit}(t_0)} \, \frac{4 \pi G}{5 c^5} \,M\, T_{\rm rh} \,  \omega_0  \,  \times  \nonumber  \\
&   \int \int \int_{\cal S}
   \left(\alpha^4+\beta^4+\gamma^4
 -{\alpha}^2\beta^2     -\alpha^2\gamma^2  -\beta^2\gamma^2\right)    \times \nonumber \\
    & \left|\text{Ei}\left[\frac13, i\omega_0  (1+z_{\rm e}) t_k \right]- \left (\frac{t_{\rm col}}{t_k} \right )^{2/3} \text{Ei}\left[\frac13, i\omega_0(1+z_{\rm e}) t_{\rm col}\right] \right|^2 \,
  {\cal F}_\text{D} (\alpha, \beta, \gamma, \sigma) \,.
\end{align}   
 In the above expression we took $(t_1, t_2)=(t_{k}, t_{\rm col})$ and  the redshift parameter at the time of collapse  $1+z_{\rm e}=1+z(t_{\rm col})$. 
For benchmark values $M=10^{22}$  g, $T_{\rm rh}=10^3$ GeV  at the frequency value  $\omega_0/2\pi=1$ Hz the coefficient
in front of the integral
is $4 \times 10^{-6}$. 

The expression (\ref{result_OmZ}) has been derived following closely the methodology established in \cite{Dalianis:2020gup}. In that work the results were presented 
for the time interval $(t_1, t_2)=(t_{\rm max},t_{\rm col})$ and  after
making a simplification, for the sake of numerical efficiency,  to approximate  $\delta_{\rm L}$ with the dispersion $\sigma$ in the integrand. 
That choice  
gave spectra close to the exact ones  for large $\sigma$ values.  
The  expression  (\ref{result_OmZ}) is exact and  it is written for
 $(t_1, t_2)=(t_{k},t_{\rm col})$. 
The overall sensitivity of the GW amplitude on the $\sigma $ value is found to be $\Omega_{\rm GW}{(f_0, t_0)} \propto \sigma^{7/2}$. In terms of the power spectrum of the curvature perturbations, ${\cal P_R}(k) \equiv A_{\rm s} \sim \sigma^2$, where a peak around $k$ is assumed, the scaling reads $\Omega_{\rm GW}{(f_0, t_0)} \propto A^{7/4}_{\rm s}$. We note that our analytic result, derived within the Zel'dovich and quadrupole approximation, has a scaling  in agreement with that reported in  ref. \cite{Fernandez:2023ddy}.  However a comprehensive comparison is not straightforward since our result originates from a monochromatic source and not from a $k$-band of a scale invariant power spectrum ${\cal P_R}(k)$. 

\subsection{GWs computed from $N$-body simulations}

For times beyond  the stage of pancake collapse $t\gtrsim t_\text{col}$  the use of  $N$-body simulation 
is  necessary  for the qualitative and quantitative description of the evolution of the overdensity. 
But also for times $t \leq t_\text{col}$ the validity of the analytic result (\ref{FT2}) gradually attenuates. 
 Utilizing an $N$-body simulation  we proceed beyond the Zel'dovich approximation
  and examine 
  the evolution of the  collapsing matter distribution.  
According to the analytic expressions (\ref{result_Pd}) and (\ref{result_Om}), the numerical tracking of the  quadrupoles 
 is the essential objective for the computation of the GW emission.
The finding of $\dddot{Q}_{ij}$  makes possible the evaluation of the total amplitude and spectrum of the GW field produced during the entire stage of the gravitational collapse.

\subsubsection{The evolution of the quadrupole}

\begin{figure}
\begin{subfigure}{.5\textwidth}
  \centering
  \includegraphics[width=1. \linewidth]{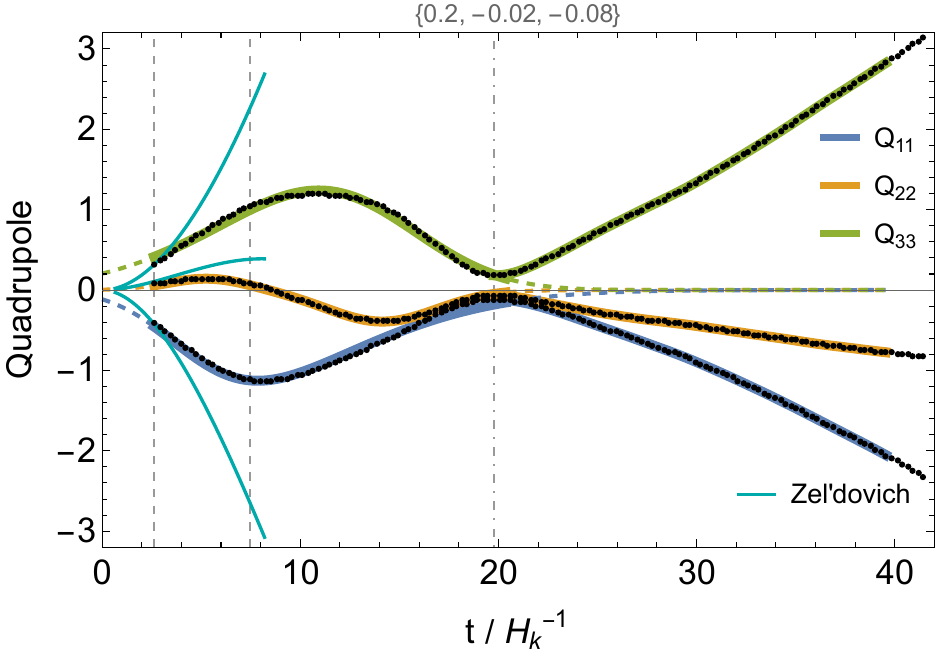}
\end{subfigure}  \quad\quad
  \begin{subfigure}{.5\textwidth}
  \centering
  \includegraphics[width=1.\linewidth]{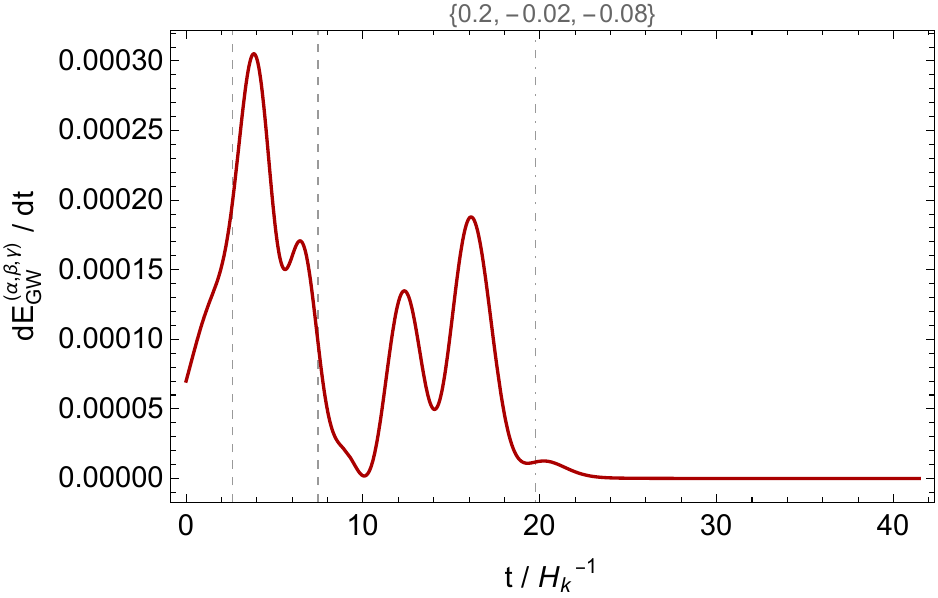}
\end{subfigure}
 \begin{subfigure}{.5\textwidth}
  \centering
  \includegraphics[width=1.01 \linewidth]{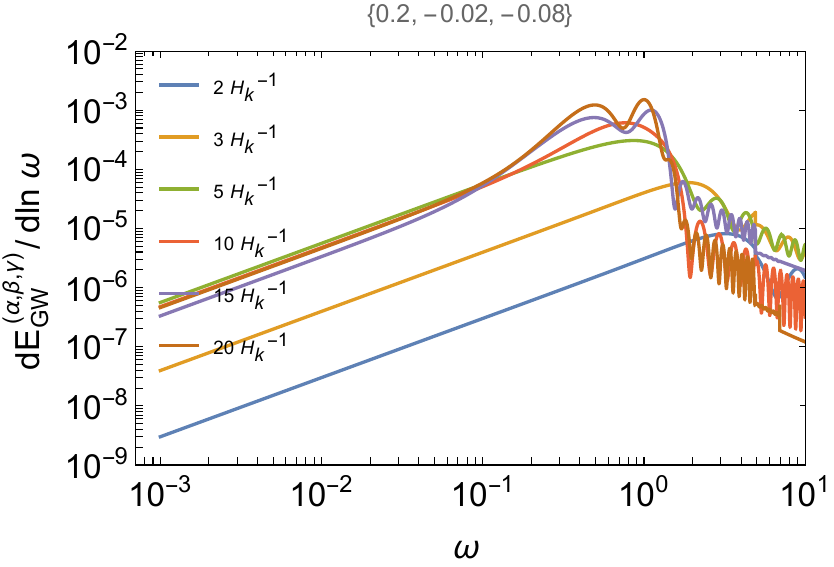}
\end{subfigure}  \quad\quad\quad
  \begin{subfigure}{.48\textwidth}
  \centering
  \includegraphics[width=1.01\linewidth]{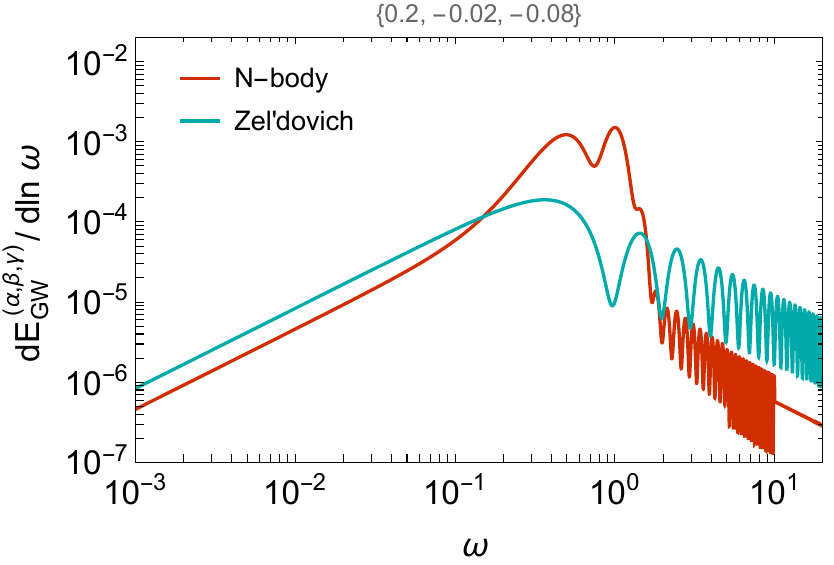}
\end{subfigure}

 \caption{\label{fonep1}~~  $N$-body results are displayed for an overdensity with deformation parameters $(\alpha, \beta, \gamma)=(0.2, -0.02, -0.08)$  simulated with $N=2500$ particles.
{\it Top left panel}: The evolution of the diagonal components of the quadrupole tensor are depicted with black dots. The fitting curves are depicted with blue, orange and green lines. The Zel'dovich solution is depicted in cyan lines. The gridlines from left to right denote the moments of turnaround, pancake collapse and bottleneck stage, in  the first two plots. 
{\it Top  right panel}: The time evolution of the GW power emitted from the Hubble patch.
{\it Bottom left panel}: The spectrum of GWs for different ending $t_{\rm end}$ moments.
{\it Bottom right panel}: Comparison of the spectra of GWs  found within Zel'dovich approximation until the moment of pancake collapse and from $N$-body simulation until virialization.
}
\end{figure}

\begin{figure}
\begin{subfigure}{.5\textwidth}
  \centering
  \includegraphics[width=1.0 \linewidth]{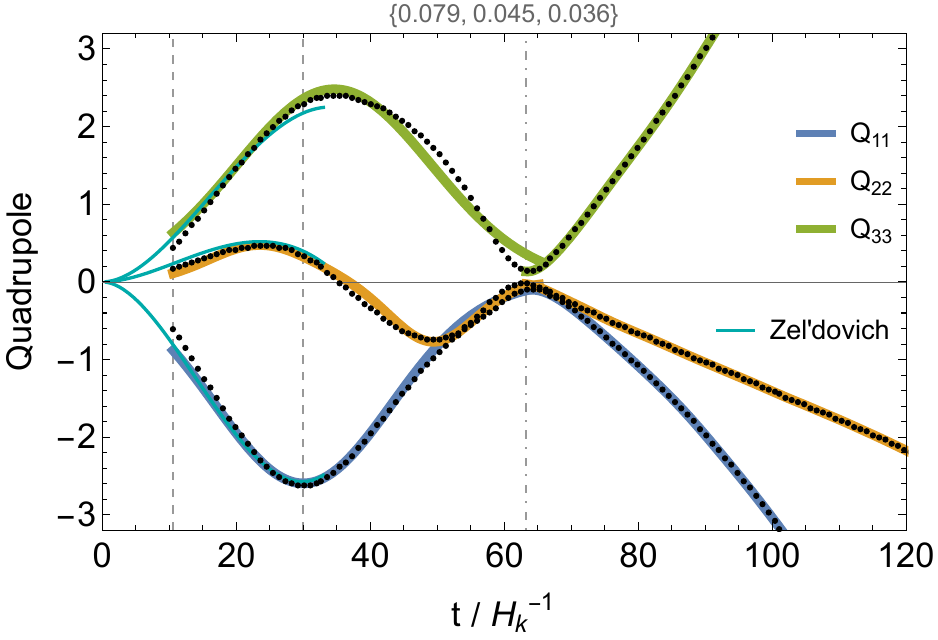}
\end{subfigure}  \quad\quad
  \begin{subfigure}{.5\textwidth}
  \centering
  \includegraphics[width=1\linewidth]{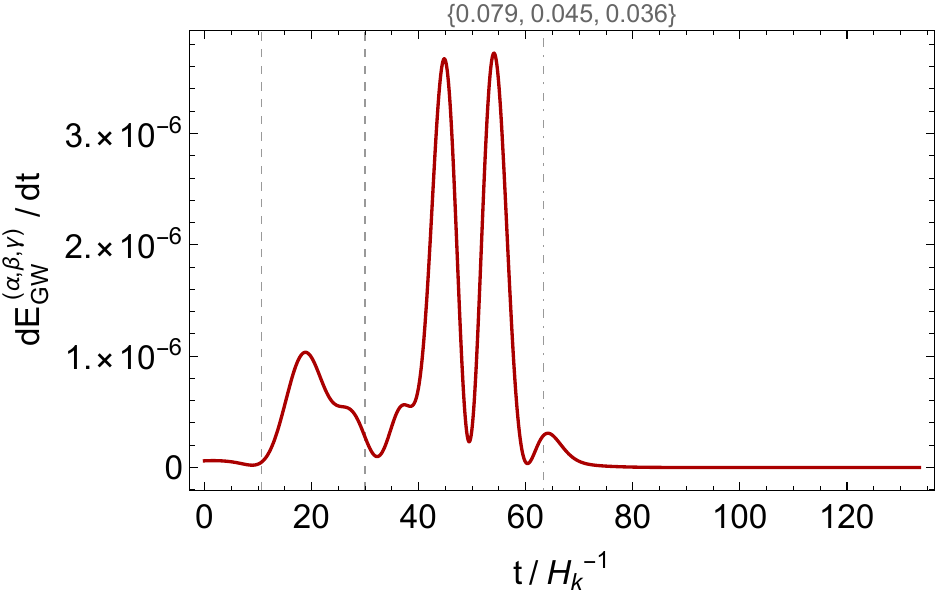}
\end{subfigure}
 \begin{subfigure}{.5\textwidth}
  \centering
  \includegraphics[width=1.0 \linewidth]{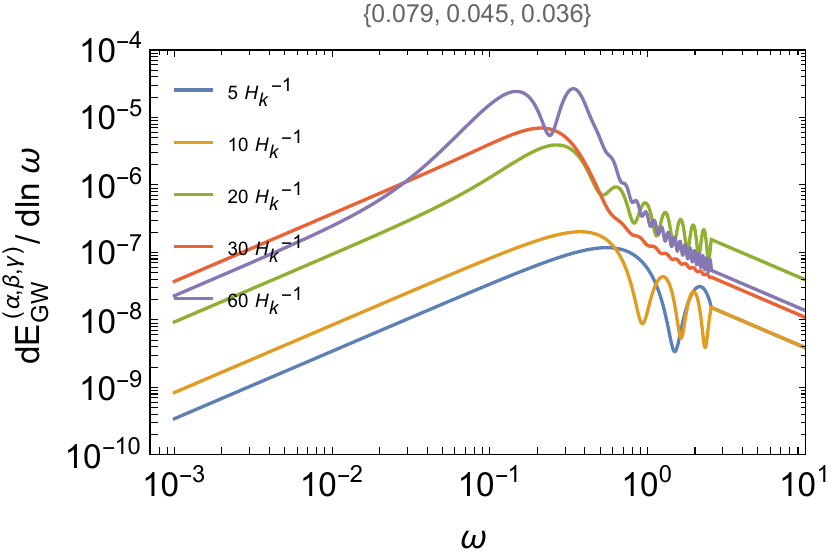}
\end{subfigure}  \quad\quad\quad
  \begin{subfigure}{.48\textwidth}
  \centering
  \includegraphics[width=1\linewidth]{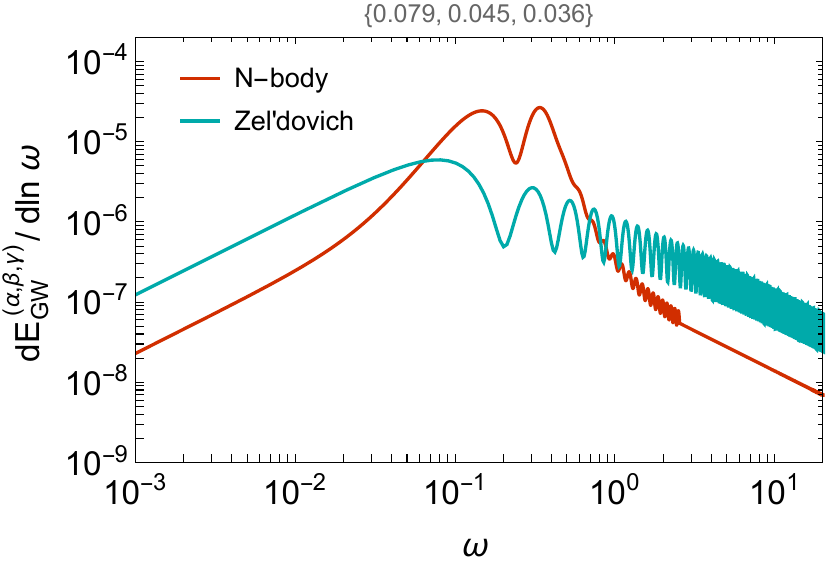}
\end{subfigure}

 \caption{\label{fonep2}~~  
As in fig. \ref{fonep1} for 
deformation parameters 
$(\alpha, \beta, \gamma)=(0.079, 0.045, 0.036)$. 
}
\end{figure}
\begin{figure}[!htbp]
\begin{subfigure}{.5\textwidth}
  \centering
  \includegraphics[width=1.0 \linewidth]{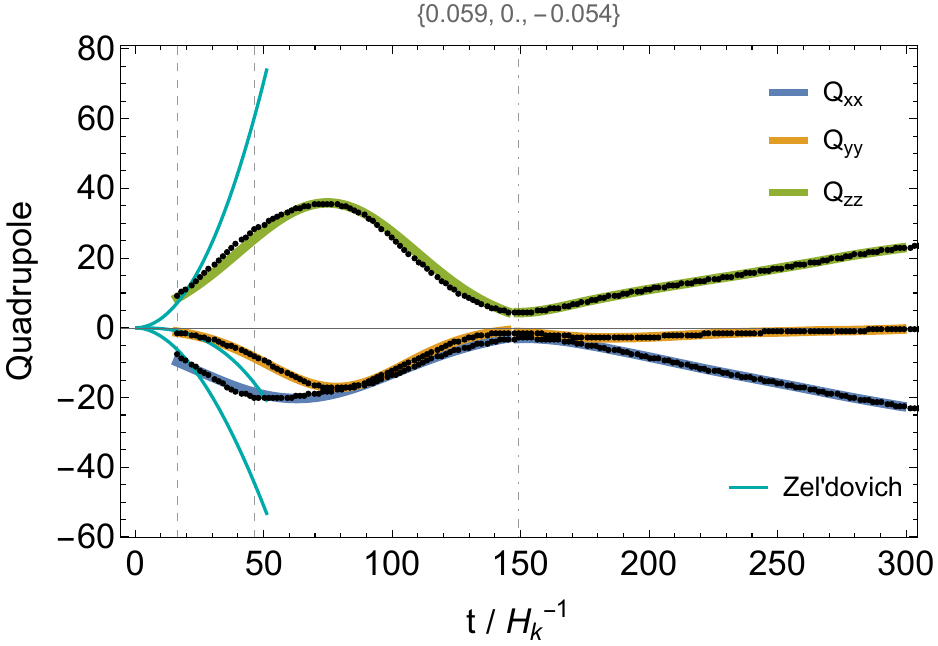}
\end{subfigure}  \quad\quad
  \begin{subfigure}{.5\textwidth}
  \centering
  \includegraphics[width=1\linewidth]{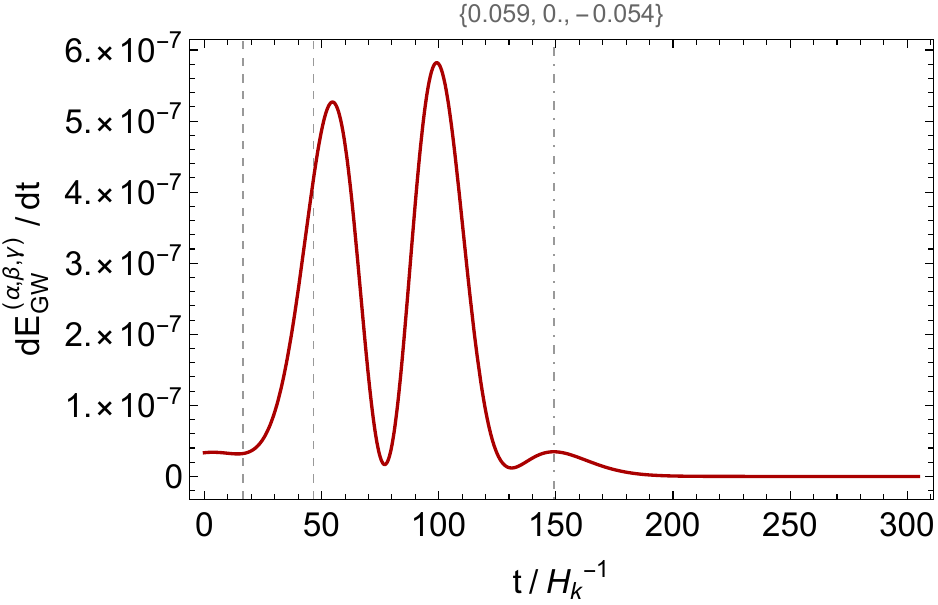}
\end{subfigure}
 \begin{subfigure}{.5\textwidth}
  \centering
  \includegraphics[width=1.0 \linewidth]{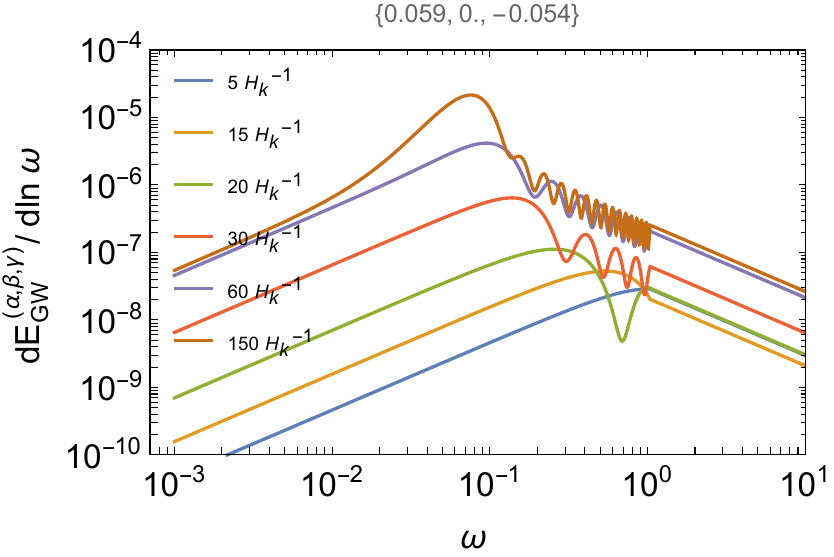}
\end{subfigure}  \quad\quad\quad
  \begin{subfigure}{.48\textwidth}
  \centering
  \includegraphics[width=1\linewidth]{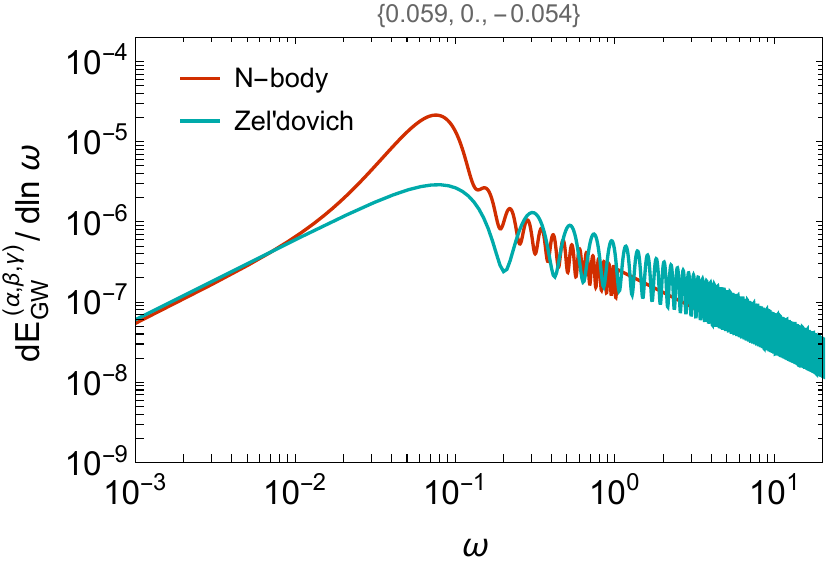}
\end{subfigure}

 \caption{\label{fonep3}~~  
 As in fig. \ref{fonep1} for 
 deformation parameters 
$(\alpha, \beta, \gamma)=(0.059, 0, -0.054)$. 
}
\end{figure}

Gravitational radiation  is produced  during collapse  with power given by eq. (\ref{Pdt}).  
The critical quantity is the quadrupole (\ref{Qij})  and the moment of inertia for the $N$-body distribution which is given by the expression
\begin{equation}
I_{ij}=\sum_{n=1}^N  m_n\,\left( |\vec{x}_n|^2 \delta_{ij}-x_i^{(n)} x_j^{(n)}  \right)\,.
\end{equation}
For  specific $(\alpha, \beta, \gamma)$  values the quadrupole is successively measured in discrete moments $t_i$ in units of $H_k^{-1}$ separated by a step $\Delta t_{\rm step}^{\rm NB} $,  small enough to give  sufficient resolution and track the subtle changes over time. 
The change of the quadrupole and the GW generation is displayed in figs. \ref{fonep1} - \ref{fonep3} for the three configurations presented in fig. \ref{fig: Vir}.
Of special interest is the fact that the quadrupole evolution of these configurations gives off efficiently gravitational radiation. 
They represent three different types of quardupole evolution which produce different shapes for the GW spectrum with a notably, strong enhancement compared to the Zel' dovich result.
However, they also share common features, generic for all configurations,  that we describe next.
In the top left panels  the result of the $N$-body simulations for the diagonal $Q_{ij}(t)$ values are depicted  with black dots.
We note that  non-diagonal components of the quadrupole arise and are taken into account, but are not displayed because they are generally about two orders of magnitude smaller than the diagonal ones.  Exception are the cases of   nearly spherical collapse where the quadrupole components  become comparable.

Apart from the distinct stages of turnaroud and pancake collapse, particular phases of the quadrupole evolution can be read off the graphical form of the $Q_{ij}(t)$ discrete data:

\begin{enumerate}

\item The first phase of the quadrupole evolution is characterized by 
a skewed bell-shaped curve 
  that takes positive or negative values along the time axis $t> t_k$. The diagonal terms are $Q_{ii} \sim M (2r_i^2-r_j^2-r_k^2)$ with $i\neq j \neq k$ and the hierarchy of $(\alpha, \beta, \gamma)$  implies that $Q_{11}$ is generally negative, 
 $Q_{33}$ is generally positive, while the sign of $Q_{22}$ is contingent on how small or large the $\beta$ parameter is with respect to $\alpha$ and $\gamma$.
An oscillation between positive and negative values
is observed in the $Q_{22}(t)$  component for a particular combination of  $(\alpha, \beta, \gamma)$ parameters.  
During the first stage of the quadrupole evolution the turnaround, the pancake collapse and shell crossing processes take place in succession. The latter process 
might  cause a flip of sign also for $Q_{11}$ and $Q_{33}$ components, which is visible in particular configurations. 

\item   The second phase of the quadrupole evolution is characterized by a bottleneck-like shape.
During this phase,  indicated by the moment $t_{\rm virial}$,  a second collapse occurs and the distribution settles into  the virialized state.  
 It is characterized by a minimum  value for each $|Q_{ii}(t)|$ diagonal component so that 
the three of them together form a bottleneck shape that corresponds to a time of maximal contraction.  The ratio $T / |U|$ maximizes before it reaches the virial value 1/2, as displayed in fig. \ref{fig: Vir}. 
The shapshot of the distribution at the moment $t_{\rm virial}$ is  displayed in fig. \ref{fanim}.

 \item The third phase of the quadrupole evolution is characterized by a roughly  monotonic   evolution.  This  is the final  phase of virialization where the ratio of the total kinetic and potential energies are stabilized to the value $K\approx- U/2$. During this stage  the $|Q_{ii}(t)|$ values increase steadily because of  the moment of inertia carried away by the  escaping particles. It is the stage where the GW emission shuts down. 

\end{enumerate}

\subsubsection{The computation of the emitted GW power}

The computation of the GW power requires  finding   the third time derivatives of  the quadrupoles.
 Successive application of numerical derivatives on the discrete data  generally introduces artificial noise. 
The method we follow to tackle this issue and avoid a possible overestimation is to fit the  quadrupole data  with suitable functions.
In particular, the quadrupole data   that span the time interval   from  $t_\text{max}$ until the  bottleneck stage, are fitted sufficiently well with a skew-normal distribution
\begin{equation}
Q_{ii}(t) \approx \frac{1}{\sqrt{2} \pi \hat{\delta}_{i}} e^{-\frac{(t - \hat{\mu}_{i})^2}{2 \hat{\delta}^2_{i}}} \,
  {\rm Erfc}\left[-\frac{\hat{b}_{i} (t - \hat{\mu}_{i})}{\sqrt{2} \hat{\delta}_{i}} \right]\,,
\end{equation}
with mean value $\hat{\mu}+ \sqrt{\frac{2}{\pi}} \,\hat{b} \hat{\delta}/\sqrt{1+\hat{b}^2}$ 
and variance $\hat{\delta}^2(1-2 \hat{b}^2/(\pi(1+\hat{b}^2)))$. This kind of fitting functions appears to be appropriate for the  diagonal quadrupole terms, which are the dominant terms of the tensor $Q_{ij}$.  For the non-diagonal terms, which generally give a subleading contribution,  we apply  a polynomial fitting. 
 Polynomial functions are also applied to fit  the data during the last stage, where virialization has been established. 
 Overall, the compound fitting allows a smooth and reliable evaluation 
of $\dddot{Q}_{ij}(t)$ from the time of entry $t_{k}$ until the end of the  virialization stage $t_{\rm virial}$. The fitting functions are depicted with thick  blue, orange and green lines over the discrete data for the $Q_{11}$, $Q_{22}$ and $Q_{33}$ components respectively and displayed in the top left panel of figs. \ref{fonep1} - \ref{fonep3}.
For comparison the quadrupole derived from the Zel'dovich solution is also displayed truncated a $t \sim t_{\rm col}$. 
We see  that, in general, a firm analytic description of $Q_{ij}(t)$ within the Zel'dovich approximation exists only for times well before $t_\text{col}$.

 The gravitational power emitted $\dot{E}_e \sim |\dddot{Q}(t)|^2$ from a Hubble patch that encloses a density perturbation   
 exhibits a very different growth profile  for different deformation parameters  $(\alpha, \beta, \gamma)$ and $\sigma$.
 In the right top panels of figs. \ref{fonep1} - \ref{fonep3}, we see that 
  pumps of  GW energy  density take  place at different moments for different configurations from  the initial time of turnaround, the stage of pancake collapse and until the establishment of virialization. 
 After the pancake collapse, the system gradually faster or slower, transforms into a virialized state with a  spherical core.
 The  virialization procedure involves geometrical changes  that might be abrupt with non-axisymmetric mass flow producing a strong GW emission.
 This is especially evident when the geometry of the bulk mass of the distribution changes from oblate to prolate-like shapes and vice versa.
 Such changes are manifest from the evolution of the quadrupole component $Q_{22}(t)\sim M (2r_2^2-r_1^2-r_3^2)$ which    fluctuates  about the zero value for particular configurations, with the positive (negative) $Q_{22}$ values  corresponding to mostly oblate (prolate) shapes, see the top left panel of figs. \ref{fonep1} and \ref{fonep2}.  A prolate-oblate geometrical change of the mass distribution  is very well 
 captured by our  fitting procedure and yields an enhanced GW radiation. 
In the latter example the quadrupole change is rather acute and  a very strong GW emission takes place. 
 When no oscillatory pattern is observed for the quadrupole components the total  amount of the GW production is  found to be  of the same order of magnitude  with the GWs produced within the ($t_k, t_{\rm col}$)  time interval. 
For  configurations with symmetric bell-shaped graphical forms for all the $Q_{ii}(t)$ components,
the power of the gravitational radiation 
is characterized roughly by a unimodal  distribution as displayed, for example, 
  in the top left panel of fig. \ref{fonep3}.  

Having specified  the entire  time evolution of the quadrupole the computation of the spectral density of the gravitational radiation is carried out.
The Fourier transform   of the third time derivative of the quadrupole,
 \begin{equation}
 \tilde{Q}^{(3)}_{ij}(\omega) =\frac{1}{2\pi} \int \dddot{Q}_{ij}(t) e^{-i\omega t}  dt \,
 \end{equation}
gives the spectral GW power 
 after using the identity $|\tilde{Q}^{(3)}_{ij}(\omega)|^2= \omega^6 |{Q}_{ij}(\omega)|^2$.
We implicitly assume a non-vanishing quadrupole only for times between the horizon entry of the perturbation and   the reheating of the Universe, i.e. in the time interval  $t_k<t<t_{\rm reh}$.  
The time of reheating is a free parameter and affects essentially  the amplitude and the spectrum of the final GW signal.
 An early 
 reheating would imply a GW signal produced only from the very first stages of gravitational collapse,  while a slow reheating implies that gravitational radiation  is produced throughout the  collapsing  and clustering process. Hence, for $t_\text{vir} < t_\text{reh}$ the final GW signal
  is sensitive to   the entire  dynamics involved in the virialization of the distribution. 
 To make the dependence on the reheating time manifest,  we compute the final GW signal for different time intervals  $[t_k, t_n]$  after parametrizing the reheating time as $t_{\rm reh}\sim t_n=  n H^{-1}_k$ 
  for positive integers $n$.
  In  figs. \ref{fonep1}-\ref{fonep3} the gradual increase of the amplitude of the spectrum $dE_{\rm GW}/d\ln \omega $ is displayed for increasing $t_n$ values.

In  the last panel of figs. \ref{fonep1}-\ref{fonep3} we display 
the $N$-body result for the GW power emitted for $t_{\rm vir} < t_{\rm reh}$ 
together with the GW power computed from the  Zel'dovich approximation  in the time interval $[t_k, t_{\rm col}]$.
Apparently the entire GW amplitude is larger because of the inclusion of the virialization.
 Violent relaxation processes that drive the system to a virialized state contribute to $\dddot{Q}(t)$ and  the GW emission  is enhanced, occasionally  significantly, given that   
  reheating occurs after the second collapse or, equivalently, after 
 the bottleneck stage.
For reheating times after the establishment of the virialization 
the emission power  practically  remains unchanged with the different 
 reheating times affecting only  the redshift  that the propagating GWs experience. 
 
 We note that an early reheating could change the composition of the system and  the evolution of the quadrupoles.
To avoid the complexities of interacting fluid dynamics  we
restrict ourselves to the discussion of the dynamics of collisionless matter. When a final state of  a virialized  spherical  distribution is reached  we expect that the reheating leaves the predictions for the GW production unmodified.

\section{The average  GW signal from gravitational collapse} \label{sec: average}

\subsection{Probability sampling of configurations }

The expectation value of the GW signal is the 
 weighted infinite sum of 
 configurations that continuously span the parameter space of $(\alpha, \beta, \gamma )$,  that is an integral  over the domain ${\cal S}$ given by eq. (\ref{result_Om}).
$N$-body simulations produce discrete results  which correspond to particular configurations. 
 Hence,  we  estimate the integral (\ref{result_Om}) numerically applying a probability sampling. 
 We divide each interval in $\alpha$, $\beta$ and $\gamma$ directions 
in regular sub-interval pieces and appropriately sum the pieces together.
The accuracy of the approximation of the breaking of the integral into a sum of pieces 
depends on 
the size of the interval considered and the number of steps taken to cross this interval. 
An adequately large size of the interval is indicated by the probability density function ${\cal F}_\text{D}(\alpha, \beta, \gamma)$ which decreases exponentially to minute values for $\alpha \gg \sigma_3$.  The  size of the step $\Delta\alpha$ has to be small enough $\Delta\alpha \ll \sigma_3$ to achieve a  sufficient resolution and not surpass configurations that might yield a strong signal. Nonetheless,  the size of the step cannot be taken too small because of computational limitations in time and resources.

\begin{table}[t]
\begin{tabular}{ |p{2.1cm}||p{1.82cm}| p{1.82cm} p{1.82cm} p{1.82cm} p{1.82cm} p{1.8cm}|  } 
 \hline
 & Integration  & $\alpha\leq \sigma_3$  &$\alpha\leq 2\sigma_3$ & $\alpha\leq 3\sigma_3$ $\quad$  & $\alpha\leq 4\sigma_3$ & $\alpha\leq 5\sigma_3$  \\
 \hline
 \hline
 $I_\text{D}$    &  1/2     & 0.05 &   0.34  &  0.46  & 0.49 & 0.50\\
$ \left\langle  \delta^2_{\rm L} \right\rangle /\sigma^2 $
 & 1  &  0.01  & 0.35   &  0.85  &  0.98   &  0.99  \\
 $\left\langle t_\text{max}\right\rangle \frac{\sigma^{3/2}}{t_k} $ & -  & 0.08  & 0.26  &  0.29 & 0.31 & 0.32 \\
 \hline
 \hline
$N_{\rm conf}$  &   $\infty$  & 78   & 650  & 2191 & 5267 & 10395 \\
 \hline 
\end{tabular}
\caption{
The values of the $I_\text{D}$, $\left \langle  a \right\rangle$ and $\sigma^{3/2}\left \langle  t_\text{max} \right\rangle/t_k$  are listed obtained after integration  over the parameter space  ${\cal S}$ and after discrete summation for $\sigma_3^2={2\times10^{-3}}$ (or  $\sigma=0.1$).  The summation results quoted are for five intervals $\alpha/\sigma_3\leq 1,2,3,4,5$ all discretized by a common step $0.2\sigma_3$. The corresponding number of configurations $N_{\rm conf}$ is also listed. 
 }
\label{table:1}
\end{table}
\begin{table}[t]
\begin{tabular}{ |p{2.1cm}||p{1.82cm}| p{1.82cm} p{1.82cm} p{1.82cm} p{1.82cm} p{1.8cm}|  } 
 \hline
 & Integration  & $\alpha\leq \sigma_3$  &$\alpha\leq 2\sigma_3$ & $\alpha\leq 3\sigma_3$ $\quad$  & $\alpha\leq 4\sigma_3$ & $\alpha\leq 5\sigma_3$  \\
 \hline
 \hline
 $I_\text{D}$    &  1/2     & 0.05 &   0.35  &  0.49  & 0.49 & 0.50\\
$ \left\langle  \delta^2_{\rm L} \right\rangle /\sigma^2 $
 & 1  &  0.01  & 0.35   &  0.85  &  0.98   &  0.99  \\
 $\left\langle t_\text{max}\right\rangle \frac{\sigma^{3/2}}{t_k} $ & -  & 0.08  & 0.26  &  0.30 & 0.31 & 0.31 \\
 \hline
 \hline
$N_{\rm conf}$  &   $\infty$  & 80   & 652  & 2277 & 5292 & 10550 \\
 \hline 
\end{tabular}
\caption{
As in table \ref{table:1} but  for a variance ten times smaller  $\sigma^2_3={2\times10^{-4}}$  (or $\sigma\approx 0.032$).
 }
\label{table:2}
\end{table}

A ruler for a reliable choice of the interval and step sizes for the discrete sum 
can be found  from the expectation values of  quantities that can be precisely  computed via straightforward integration. We first consider three pilot examples: the volume $I_{\rm D}$  of the Doroshkevich PDF in the interval ${\cal S}$ (that we referred in sec. \ref{subsec: Dor}), the expectation values of the sum deformation parameter $(\alpha+\beta+\gamma)^2=\delta^2_{\rm L}$ (\ref{vevdelta}) and the turnaround time $t_\text{max}$,
\begin{align}
 \left \langle t_\text{max} \right\rangle  = \,{t_k} \,  \,  \int_{0}^{1/2}\,d\alpha \int_{-\infty}^{\alpha}d\beta \int_{-\infty}^{\beta} d\gamma
\, 
\,d\alpha d\beta d\gamma\, 
 {\cal F}_\text{D} (\alpha, \beta,\gamma, \sigma_3)
\left( \frac{1}{2\alpha } \right)^{3/2}\,.
  \label{texp}
\end{align} 
In tables \ref{table:1} and \ref{table:2} we list the exact  and five approximate values  for 
$I_\text{D}$, $\left\langle \delta^2_{\rm L} \right\rangle $ and $\left \langle t_\text{max} \right\rangle$ 
given  by integration 
and  by a discretized summation respectively.
The integration interval is the entire parameter space ${\cal S}$ while for the discretized summation we consider five different intervals for the positive parameter $\alpha$ which are: $\alpha \leq \sigma_3$, $\alpha \leq 2\sigma_3$, $\alpha \leq 3\sigma_3$, $\alpha  \leq 4 \sigma_3$  and $\alpha \leq 5 \sigma_3$. For each $\alpha$-interval the corresponding  $\beta$- and $\gamma$-intervals are dictated  
by  the bounds  (\ref{hiera}).
The step-size that spans the three-dimensional interval is $\Delta \alpha=\Delta\beta=\Delta \gamma =0.2 \sigma_3$, which is a choice found to give a sufficient resolution.
In table \ref{table:1} we take  $\sigma_3^2=2\times 10^{-3}$ (or $\sigma=0.1$) for the variance of the Doroshkevich PDF. In table \ref{table:2} we take $\sigma_3^2=2\times10^{-4}$ (or $\sigma\simeq 0.032$).
The values listed in the tables show that the discretized summation
 over the interval $\alpha \leq 4\sigma_3$ approximates very well the exact values. 
 The corresponding number of configurations are $N_\text{conf}=5267$ for $\sigma=0.1$ and $N_\text{conf}=5329$ for $\sigma=0.032$.  Decreasing the size of the step to $\Delta\alpha=0.1$ would increase $N_{\rm conf}$ an order of magnitude without significantly improving the precision of the  results.

Having specified a reliable choice for the sizes of the interval and the step, i.e. $\alpha \leq 4 \sigma_3$ and $\Delta\alpha=\Delta\beta=\Delta\gamma=0.2\sigma_3$ respectively, we proceed with the calculation of the power of the  gravitational radiation produced. 
In this respect we approximate   the integral with a discretized summation
 \begin{align} \nonumber
\left\langle \frac{dE_{\rm GW}}{d\ln \omega}\right\rangle & =\int\int\int_{\cal S}  
\,d\alpha d\beta d\gamma\,  
  \frac{4 \pi G}{5 c^5}  \omega |\tilde{Q}^{(3)}_{ij}\left(\omega)\right)|^2 
 {{\cal F}_\text{D} (\alpha, \beta,\gamma, \sigma_3)}    \\
& \approx   \frac{4 \pi G}{5 c^5}  \omega  
\sum_{\alpha_i} 
\sum_{\beta_i} \sum_{\gamma_i} \Delta \alpha_i \Delta \beta_i \Delta \gamma_i\, |\tilde{Q}_{ij}^{(3)}\left(\omega, \alpha_i, \beta_i,\gamma_i, \sigma_3)\right)|^2   \,  {{\cal F}_\text{D} (\alpha_i, \beta_i,\gamma_i, \sigma_3)}\,. \label{Pdiscr} 
\end{align}  
Contrary to the examples listed in tables \ref{table:1} and \ref{table:2}
 the calculation  of (\ref{Pdiscr}) is a rather elaborated procedure. 
It involves $N_{\rm conf}$ runs for the $N$-body simulation with initial conditions $(\alpha_i, \beta_i,\gamma_i, \sigma_3)$ followed by the necessary data analysis for the finding of the quadrupole and its Fourier modes $\tilde{Q}_{ij}(\omega)$. 
The repetition of  the runs and data analysis for $N_\text{conf}$ times increases significantly the computational cost. 
For the computation-intensive purposes of the project we utilized a small  cluster of ten computers.
Restricting ourselves to a computation that takes time less than a couple of months puts a limitation to the 
number of particles in our $N$-body simulations. 
A large $N$ increases the run-time of the $N$-body simulations and, moreover, overloads the data analysis. For $N={\cal O}(10^4)$ particles and sufficiently small step $\Delta_{\rm step}^{\rm NB}$ and large $t_{\rm end}$ a single simulation takes time ${\cal O}$(10) min and the data analysis several hours.
We choose $N=2500$ after having checked that  by varying this value by a couple of times there are no qualitative changes  to the amplitude and evolution of the quadrupole.

\begin{figure} [!t] 
  \begin{subfigure}{.48\textwidth}
  \centering
  \includegraphics[width=1.0 \linewidth]{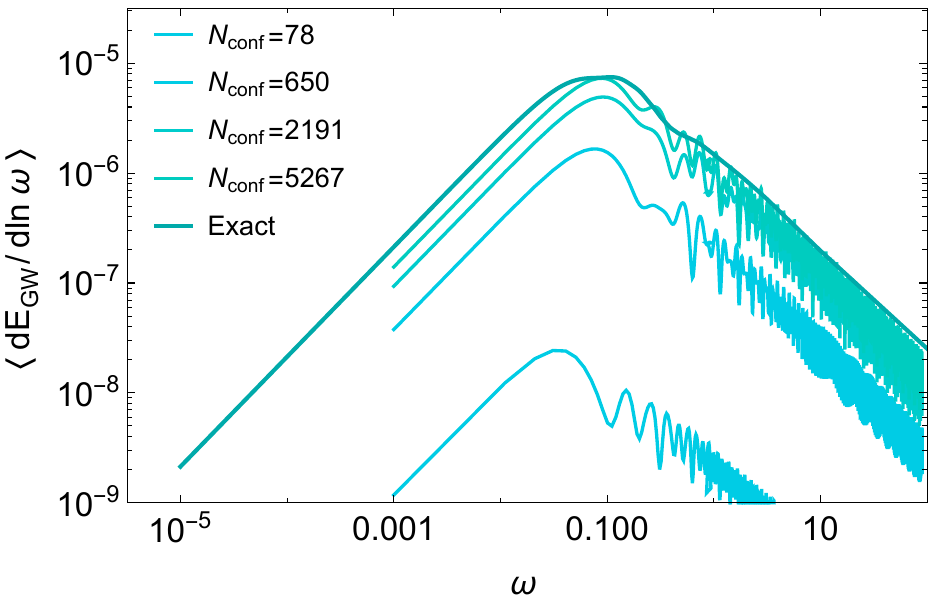}
\end{subfigure}  \quad\quad
  \begin{subfigure}{.48\textwidth}
  \centering
  \includegraphics[width=1\linewidth]{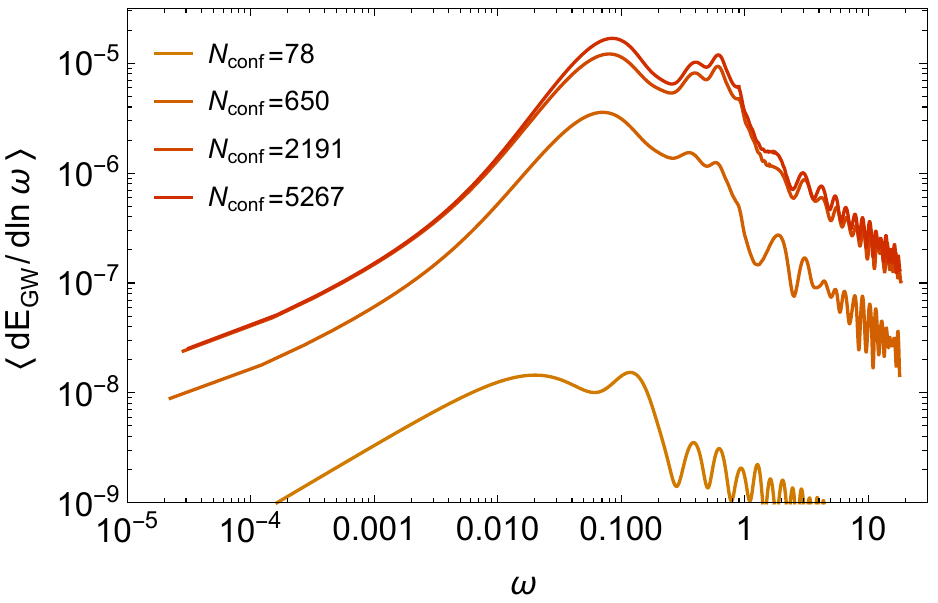}
\end{subfigure}
 \caption{\label{fcomp}~~  
{\it Left  panel}: 
The progressive convergence of the GW spectra  produced within the Zel'dovich description from a discrete summation of $N_{\rm conf}$ configuration is displayed. The exact average result   found from integration is also depicted. 
{\it Right Panel}: 
The  GW spectra produced from a discrete summation of GW spectra found after running  the  $N$-body simulation for $N_{\rm conf}$ different initial conditions are displayed. The plots are for $\sigma=0.1$.
}
\end{figure}

The average value for the GW signal found after a discrete  weighted sum of  $N_{\rm conf}\sim 5267$   different configurations for $\sigma=0.1$ 
is expected to approach very well the actual one. 
This is supported in part from tests of average  values of single-valued quantities
listed in tables \ref{table:1} and \ref{table:2}.
A second  important test 
is provided by the comparison of the Zel'dovich solution with that produced from the  discrete summation. 
The l.h.s. of eq. (\ref{Pdiscr}) can be computed exactly within the Zel'dovich approximation   (\ref{FT2}) and within the time interval $[t_k, t_{\rm col}]$. 
In the left panel of  fig. \ref{fcomp} we display the progressive convergence of the summation result and the analytic one with increasing $N_{\rm conf}$.  
For the summation and the analytic result we considered configurations that   experience a turnaround at times $t_{\rm max} < 10^2 t_k$ that is for  $\alpha>0.023$. 
The importance of this test is that it  exhibits the convergence of the summation procedure  (r.h.s. of eq. (\ref{Pdiscr})) at the exact result (l.h.s. of eq. (\ref{Pdiscr})) for a continuous range of values that span five orders of magnitude in angular frequencies.
We note that for smaller $\alpha$ values, $\alpha\lesssim 0.02$ the analytic result would differ in the 
 infrared frequency range, as demonstrated in fig. \ref{ZeldintIR}. In this case  a step of a smaller size would be required for an accurate summation procedure.

\subsection{The average spectrum of GWs}

The aforementioned tests solidify the dependability of our methodology and we can firmly proceed to the estimation of the average power of the gravitational radiation. 
In the right panel of fig. \ref{fcomp} the average spectrum of GWs is displayed after running the $N$-body simulation for increasing $N_{\rm conf}$ number.  The displayed spectra are approximations of the spectrum of GWs emitted 
 from the moment of entry of the overdensity $t_k$ until a final moment $t_{\rm end} \gg t_{\rm col}$ that  virialization has been established. 
For the time interval $[t_k, t_{\rm end}]$ the tensor $Q_{ij}(t, \alpha_i, \beta_i, \gamma_i)$ is found and the gravitational power emitted is computed from the r.h.s. of eq. (\ref{Pdiscr}).
A minimum number of configurations is required for a solid result. 
For $N_{\rm conf}\lesssim 100$ the summation 
fails to capture the precise form and the amplitude of the spectrum  because configurations with significant contributions are omitted. For $N_{\rm conf} \gtrsim 2000$ the fine details of the GW spectrum are revealed. 
The average GW spectrum  features an additional peak after the inclusion of the virialization process.  
The first peak is shaped by the stages of turnaround, pancake collapse and the return to the nearly spherical shape. 
The peak at higher frequencies is presumably associated  with the violent relaxation processes that involve shell crossings and redistribution of the particles.
The two peak structure  in the  spectrum  appear after the proper superposition of $N_{\rm conf}$ individual spectra, each with a different peak structure, see   figs. \ref{fonep1}-\ref{fonep3} . 

\begin{figure} 
  \begin{subfigure}{.48\textwidth}
  \centering
  \includegraphics[width=1.0 \linewidth]{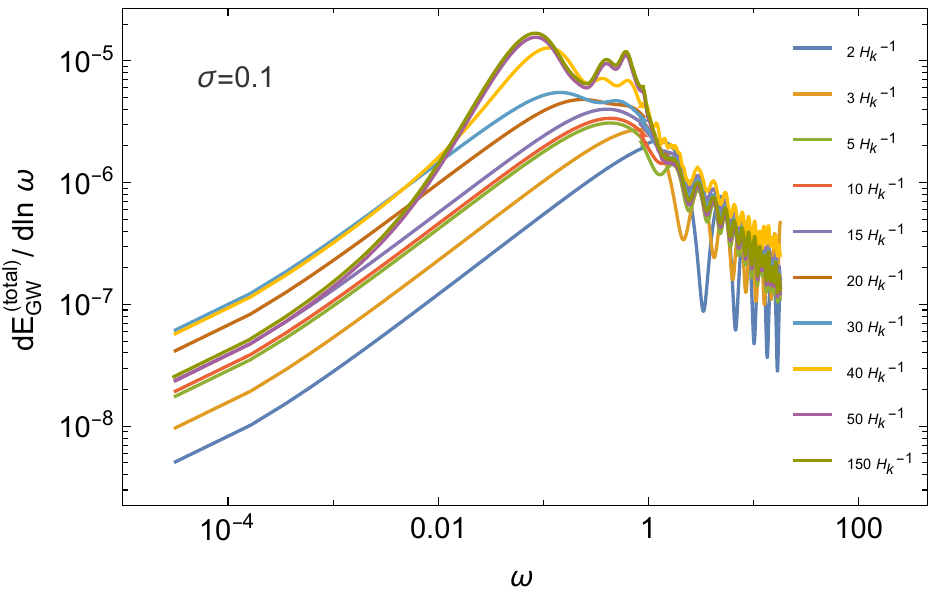}
\end{subfigure}  \quad\quad
  \begin{subfigure}{.48\textwidth}
  \centering
  \includegraphics[width=1\linewidth]{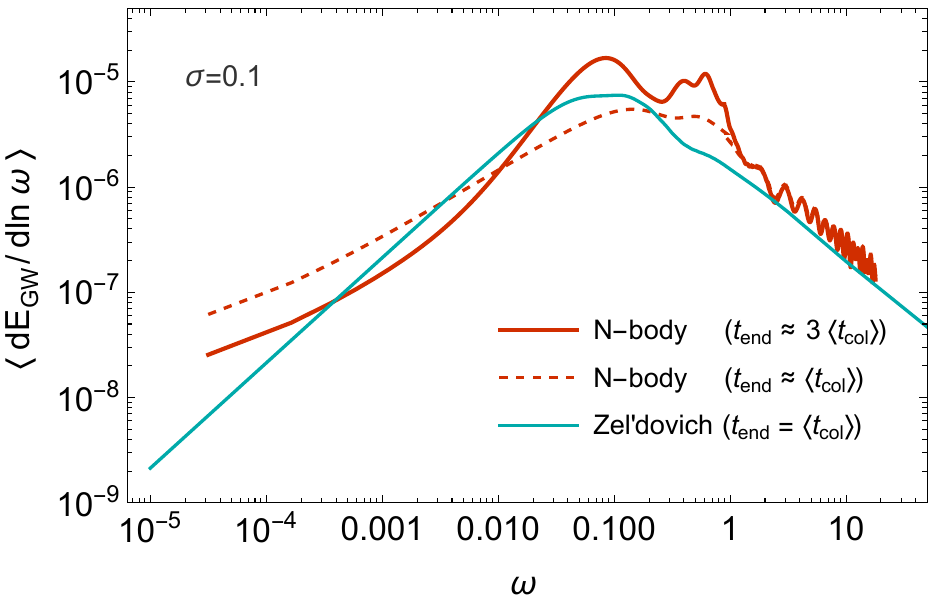}
\end{subfigure}
 \caption{\label{fig: avGWs}~~  
Results of the average GW spectrum (\ref{result_Pd}) for $\sigma=0.1$.
In the {\it left panel} we display the gradual growth of the GW spectrum as the collapsing process evolves. In the plot legends
 the corresponding  truncation time of the quadrupole evolution can be read off in Hubble time units. 
In the {\it right panel} the averaged GW spectrum for the particular times $t_{\rm end}=\left\langle  t_{\rm col} \right\rangle$ and $t_{\rm end}=3 \left\langle t_{\rm col} \right\rangle$ are depicted. For comparison the average Zel'dovich spectrum is depicted with the cyan line.
}
\end{figure}

In fig. \ref{fig: avGWs} the average GW spectrum is displayed in a time-lapse sequence.
The $Q_{ij}$ evolution is truncated at  different moments. 
The truncation reveals step by step  the effects of the virialization process.
 It can be also viewed as  reheating times under the assumption that reheating does not modify the spectrum significantly. 
Hence, the reheating time and the peak structure are correlated and can specify the reheating temperature. 
In the left panel, the lower and upper curves are respectively the spectra 
of  GWs emitted within  2 Hubble and 150 Hubble times  after the entry of the overdensity.
The gravitational radiation gradually increases as the evolution of the  gravitational system crosses the stages of pancake collapse and violent relaxation. 
We also see that after $50$ Hubble times the signal has reached its maximum value and  its multi-peaked structure has emerged.
After that time  the form of the spectrum settles into a time-invariant shape. Although there are configurations that continue to give off  GWs after $\sim 50 H_k^{-1}$ their contributions to the total spectrum is negligible. The GW signal is dominated by those configurations which evolve fast and source a sizable $\dddot{Q}_{ij}$.

In the right panel of fig. \ref{fig: avGWs} we display the average  GW spectrum at the  average moment 
of pancake collapse,
\begin{equation}
\left\langle t_{\rm col} \right\rangle = 2\sqrt{2} \left\langle t_{\rm max} \right\rangle \approx 0.9  \,\sigma^{-3/2} \,t_k  \,.
\end{equation}
It is $\left\langle t_{\rm col} \right\rangle \sim 28 H_k^{-1}$ for $\sigma=0.1$.
At the time $\left\langle t_{\rm col} \right\rangle$ the 
structure of the spectrum is characterized by a broad single peak, in quite good  agreement with the Zel'dovich result which is also displayed for comparison.  However, 
hump-like  features around the peak also appear 
which originate from  shell-crossing processes  occurring in configurations that evolve faster than the average and have  gone through the pancake collapse phase earlier on. 
Within $3\left\langle t_{\rm col} \right\rangle$ the average GW spectrum has already attained its final form.
It is characterized by a first peak at $\omega_{\rm peak} \sim H_k /10$
and a second  peak, with roughly one half the amplitude, positioned at the angular frequency $ \omega_{\rm peak} \sim H_k /2$  for $\sigma=0.1$. 

The amplitude of the spectrum first peak, found with $N$-body simulations, is given approximately by the expression,
$ \left\langle {dE_{\rm GW}}/{d\ln \omega}  \right\rangle \sim \, 2 \times 10^{-5}$ for $\sigma=0.1$. For smaller $\sigma$ values the spectrum changes according to the discussion in sec. \ref{GWzel}. 
Also, the scaling of $\left\langle {dE_{\rm GW}}/{d\ln \omega}  \right\rangle$ with frequency is $\omega^{3/2}$ at the infrared part of the spectrum and  $\omega^{-1}$ in the ultraviolet.

\subsection{The observable GW signal}

The observable GW signal is given by eq. (\ref{result_Om}).
It is determined after three input parameters are specified: the mass $M$ of the perturbations, the deviation $\sigma$ and the reheating temperature $T_{\rm rh}$. We will assume here that the time of reheating is late enough for the virialization procedure to complete, i.e. $t_{\rm rh} 
\gtrsim  3\left\langle t_{\rm col} \right\rangle$.  The amplitude of the observable spectrum about the peak is
\begin{align}\label{Omegapeak}
\left. \Omega_\text{GW}(t_0) \right|_{\rm peak}\,\sim  \, 10^{-6} \, \sigma^{7/2} \,
\left(\frac{M}{M_\odot}\right)^{2/3} 
\left(\frac{T_\text{rh}}{\text{GeV}}\right)^{4/3},
\end{align}
where $M_\odot$ the solar mass.
The spectrum features two peaks which stand out: one broad
at frequency given by the expression $f_{\rm peak} \sim 4 \times 10^{-9}  
 \left({M}/{M_\odot}  \right)^{-1/3} 
   \left({T_\text{rh}}/{ \text{GeV}} \right)^{1/3}$ Hz
 and one ruffled peak at frequencies $\sim 5 f_{\rm peak}$.
We note that  if the power spectrum of curvature perturbations ${\cal P_R}(k)$  is (nearly) scale invariant, instead of being enhanced around a particular scale $k$ with corresponding horizon mass  $M$,  the characteristic frequency is determined by the reheating temperature $T_{\rm rh}$.

The  differential energy density parameter of the stochastic GW background can  be optimistically detected by GW experiments that we describe next.
 $\Omega_{\rm GW}(t_{0}, f_0)$ is constrained from above 
from BBN/ CMB constraints \cite{Maggiore:1999vm, Caprini:2018mtu}. 
 It is also indirectly  constrained from the PBH production rate associated with the gravitational collapse, see e.g.  \cite{Carr:2020gox} for a review.
A maximal  $\sigma$ value maximizes both $\Omega_{\rm GW}$
and the PBH abundance.
We  recall that 
the PBH fractional abundance, $f_{\rm PBH} \equiv \Omega_{\rm PBH}/\Omega_{\rm DM}$,  is $f_{\rm PBH} \sim 10^{19} \gamma_{\rm M} \beta_{\rm PBH} (T_{\rm rh}/ 10^{10} {\rm GeV}) $.
$\beta_{\rm PBH}$ is the PBH  production  rate which is given by the expression  $\beta_{\rm PBH}\approx 0.056 \sigma^5$ during EMD and for large $\sigma$ values \cite{Harada:2016mhb}. 
 We note that the PBH mass does not coincide but is a  fraction of the horizon mass  $M_{\rm PBH}=\gamma_{\rm M} \,M$.  
The GW signal can be strong enough to be detectable and possibly discriminated.

\begin{figure} 
  \begin{subfigure}{.48\textwidth}
  \centering
  \includegraphics[width=1.0 \linewidth]{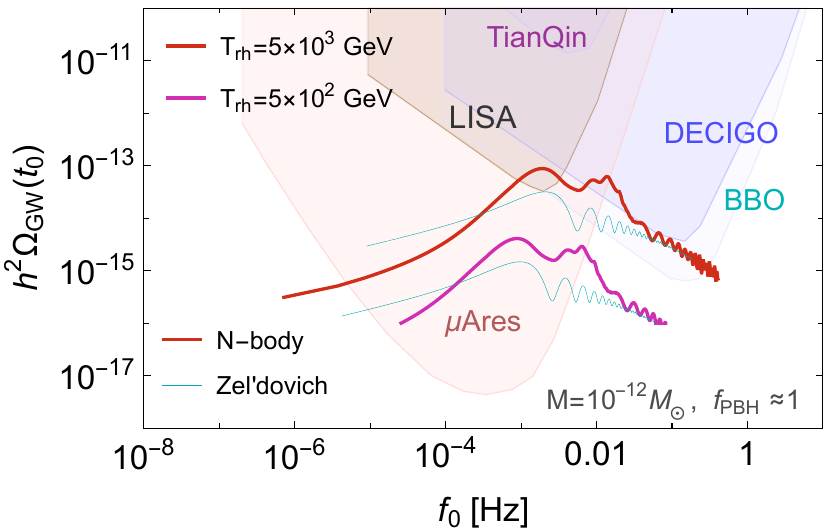}
\end{subfigure}  \quad\quad
  \begin{subfigure}{.48\textwidth}
  \centering
  \includegraphics[width=1\linewidth]{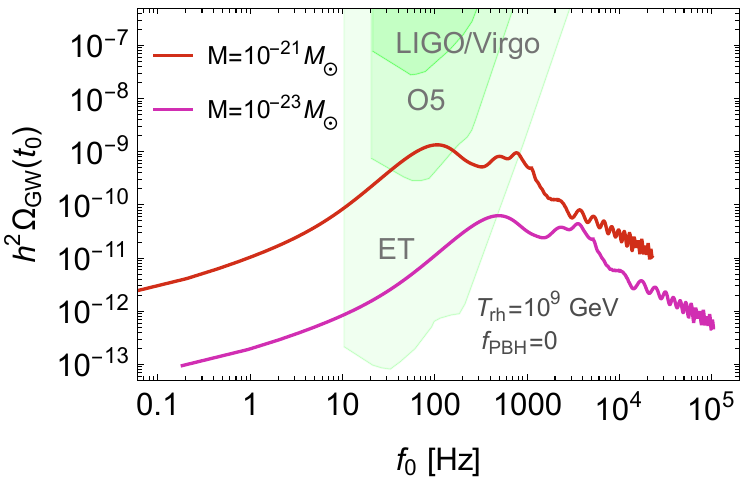}
\end{subfigure}
 \caption{\label{fig: LISALIGO}~~  
{\it Left  panel}:  GW spectra  within the  LISA-BBO frequency band that are associated with maximal PBH abundance $f_{\rm PBH}\sim 1$ and horizon mass $M=10^{-12} M_{\odot}$.
With reddish lines the $N$-body result is displayed and with cyan lines, for comparison, the Zel'dovich result. 
The upper red line  is for $\sigma = 0.01$ and $T_{\rm rh}\simeq 5\times 10^3$ GeV and the lower purple line is for
$T_{\rm rh}=5\times 10^2$ GeV. 
{\it Right Panel}:
The GWs associated with  promptly evaporating PBHs 
 for $\sigma=0.1$ and $T_{\rm rh}=10^9$ GeV. }

\end{figure}

\begin{figure} 
  \centering
  \includegraphics[width=1\linewidth]{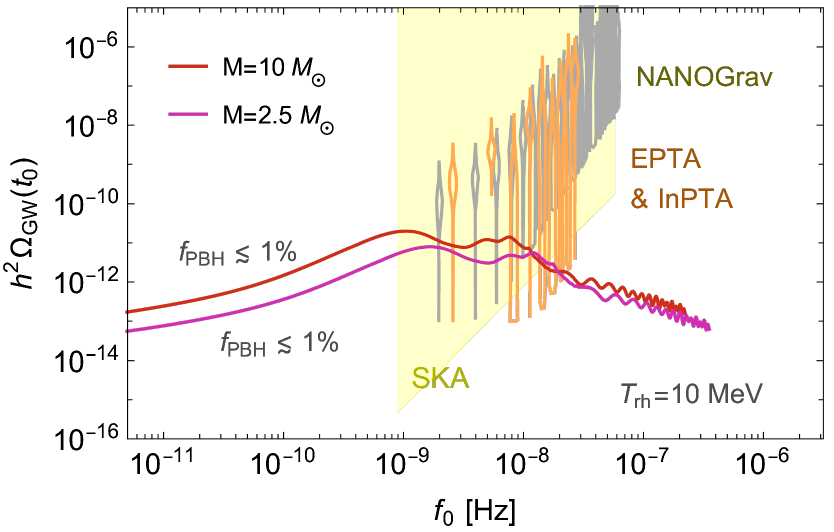}
 \caption{\label{fig: PTA}~~  
The GW spectrum in the PTA frequency range, for horizon mass $M=10 M_{\odot}$ (red line) and $M=2.5 M_{\odot}$ (purple line) and for $\sigma=0.03$.
 The GW signal is associated with a PBH population with fractional abundance $f_{\rm PBH}\approx 10^{-2} \gamma_{\rm M}$.  The reheating temperature is taken to be $10$ MeV.
The  violins in gray show the posteriors from NANOGrav \cite{NANOGrav:2023hvm} and in orange from EPTA and InPTA \cite{EPTA:2023fyk}.
}
\end{figure}

We choose three benchmark frequency bands to display the spectrum of the observed GWs. These are the regions of hertz-kilohertz, millihertz and  nanohertz where operating or designed GW experiments have an increased sensitivity.
 LIGO-Virgo-KAGRA\cite{Harry:2010zz, Acernese:2015gua, KAGRA:2020tym} and Einstein telescope \cite{Sathyaprakash:2012jk}, which  are  ground based interferometers, are
 sensitive at  $10-10^3$ Hz.  
Space-based designed interferometers LISA \cite{Audley:2017drz}, Taiji \cite{Guo:2018npi}, Tianqin \cite{Luo:2015ght}, Decigo \cite{Seto:2001qf, Sato:2017dkf}, Big Bang Observer (BBO) 
 \cite{Crowder:2005nr} and $\mu$Ares \cite{Sesana:2019vho}
are mostly sensitive at  $10^{-6} - 10^{-1}$ Hz.  
Pulsar Timing Arrays (PTAs) 
experiments 
are sensitive at $10^{-9} - 10^{-7}$ Hz.  
At this low frequency range a measurement of a stochastic GW background is reported by 
NANOGrav collaboration \cite{NANOGrav:2023gor, NANOGrav:2023hde, NANOGrav:2023hvm},  EPTA and InPTA \cite{EPTA:2023fyk, EPTA:2023sfo, EPTA:2023xxk, InternationalPulsarTimingArray:2023mzf},  PPTA \cite{Reardon:2023gzh, Zic:2023gta,  Reardon:2023zen} and CPTA \cite{Xu:2023wog}. 
These experiments detect  a stochastic GW background signal with  approximate amplitude of  around $h^2\Omega_{\rm GW} \sim 10^{-8}$ at the frequency $f\sim 10^{-8}$ Hz. 
In that frequency rang
SKA \cite{Janssen:2014dka} experiment will also operate.
In figs. \ref{fig: LISALIGO} - \ref{fig: PTA}  we display GW spectra
that can be probed by the aforementioned  experiments and
are  produced by specific examples of gravitational collapse 
that we describe next.

\begin{itemize}

\item
 In the left panel
   of fig. \ref{fig: LISALIGO} we 
focus on   density perturbations at a mass scale that can be probed by LISA experiment.
  We assume a maximal  
 $\sigma$ value which yields PBHs that saturate the dark matter density, i.e. $f_{\rm PBH} \sim 1$, and have mass $M_{\rm PBH} =10^{-12}\gamma_{\rm M}  M_{\odot}$.  We also assume $\gamma_{\rm M}\sim 0.1$ in order to maximize the GW signal (allowing smaller $\gamma_{\rm M}$ values the GW signal increases significantly).
The gravitational collapse  yields a GW signal that can be inside the LISA sensitivity curve. 
 In the same plot we consider two different reheating temperatures to stress the $T_{\rm rh}$ dependence:
  The GW spectra have  peaks at different frequency values  due to the different expansion history that they  
   experience. Hence, in the event of GW detection the reheating temperature of the Universe can be also probed. If additionally a PBH counterpart is detected  the determination of  $T_{\rm rh}$ can be a robust result. For comparison, the Zel'dovich result (\ref{result_OmZ}) is shown in the plot.
 
 \item
In the right panel of fig. \ref{fig: LISALIGO} we 
focus on the LIGO-Virgo-KAGRA frequency band.
 The corresponding mass scale is rather small, $M_{\rm PBH}/\gamma_{\rm M}\sim 10^{-20} M_{\odot} (T_{\rm rh}/10^{9}{\rm GeV})(f_0/100 {\rm Hz})^3$ and PBH  possibly produced are expected to  evaporate in the early Universe.
 An EMD era at such high energy scales is motivated by the slow reheating of the inflaton field.
 We choose $T_{\rm rh}=10^9$ GeV that is realized in scenarios where the inflaton decays gravitationally. We also
  take $\sigma=0.1$ and consider  two perturbation  mass scales: $M=3 \times10^{-21}$ and  $M=3 \times10^{-23}$. We also consider $\gamma_{\rm M} \ll 1 $ so that  PBH explosions occur fast enough to evade the BBN constraints \cite{Carr:2020gox}. Although $f_{\rm PBH}=0$ due to Hawking evaporation, PBH remnants may contribute to the DM density in the Universe, see e.g. \cite{Barrow:1992hq, Hidalgo:2011fj, Dalianis:2019asr, Dalianis:2021dbs, Domenech:2023mqk, Franciolini:2023osw}.
We see that the GW signal is well inside the  sensitivity curve of Einstein telescope, as well the advanced LIGO-Virgo.
It is interesting that LIGO-Virgo already poses constraints and upper bounds on stochastic GW background and PBHs scenarios. 
Alternatively, (weak) constraints on the reheating temperature of the Universe after inflation can be derived. 
 
 \item
Fig. \ref{fig: PTA} is about  PTA experiments. 
In that frequency range there  is accumulative evidence that a stochastic GW background exists with a significant  amplitude. 
In the plot we have placed the first Fourier bins posteriors of the common EPTA and InPTA signal, represented by the orange violin areas \cite{EPTA:2023fyk} and the NANOGrav signal in gray \cite{NANOGrav:2023hvm}.
Two GW spectra are plotted which are produced by two different overdensities: the first has mass $M=10 M_\odot$ and the second 
$M=2.5 M_\odot$ for $\sigma=0.03$. Each GW signal is associated with a PBH population with $f_{\rm PBH}\approx 10^{-2}\gamma_{\rm M}$. The horizon masses are rather large and a late reheating is necessary in this scenario. We take $T_{\rm rh}=10$ MeV. 
We see that  GWs that originate from gravitational collapse in an EMD Universe 
could contribute to the observed signal. The particular examples presented can fit better the lowest frequency bins.
Fully explaining the observed PTA signal is constrained by the bounds on the PBH abundance. 

\end{itemize}

The detection and identification of the properties and origin of a stochastic GW background is certainly challenging,  nevertheless, as we  exemplified,  
a  comprehensive understanding of the gravitational collapse can give  distinct predictions and  informative  constraints.

 \section{Conclusions} \label{sec: concl}

In this work we studied the evolution of matter anisotropic gravitational collapse and the associated GW emission via the use of $N$-body simulations.
The great advantage of $N$-body simulations is that they reveal the evolution of the mass distribution all the way to and beyond the pancake collapse stage. 
The pancake stage 
is the limit of validity for the approximate solutions found  
within the Zel'dovich framework, which  were used to compute the GW spectrum in a preceding work \cite{Dalianis:2020gup}. 
Through $N$-body  techniques,  insights into
 subsequent stages during which 
 the clustering action of gravity becomes manifest,  
are attained. 
The full evolution of perturbations
from the moment the inhomogeneity drops out of the Hubble flow until the stage of halo formation is traced.
The evolution involves anisotropic mass motions,  compression and virialization processes which source gravitational radiation. 

The total energy radiated is distributed among multipoles and we consider that  the dominant contribution
comes from the system's quadrupole moment.
Density perturbations  with different geometric profiles have a different GW emissivity with stronger gravitational radiation  arising  
from large density perturbations with acute aspherical profile. 
Depending on the initial degree of asphericity the resulting spectrum of the produced GWs might display a single peak, similar to that found  within the Zel'dovich approximation,  or display an extra peak  attributed to processes occurring during the virialization stage which   Zel'dovich approximation cannot capture since it eventually breaks down.
Each $N$-body run simulates the evolution  of a particular configurations. 
In order to find the average GW signal produced by a Hubble patch,    we perform a
survey of a large number  $N_{\rm conf} \sim 5000$ of
 configurations that span the space of the different deformation possibilities with a sufficiently small discrete step. 
 This is implemented after  $N_{\rm conf}$ runs of the $N$-body simulation   each time with appropriate initial conditions.
The resulting GW signal is a superposition of all the configurations. 
In the averaging procedure we consider the Doroshkevich probability density function, invoking a Gaussian distribution
 for the shape deformations for the  overdensities.
 The final  spectrum displays a two peak structure, with a rough second peak, 
 which reflects the dynamics of virialization processes.

In this work we postulated that a pressureless gravitational collapse has been realized in the very early Universe. A motivation for such postulation  is that the
 GW signal produced could be within the detection sensitivity region of current or designed GW experiments.  A significant PBH population associated with the GW signal could be also produced and contribute to the dark matter density. 
A detection event of such a GW signal would have, among others, two 
important consequences. First,  valuable insights into the virialization dynamics would be gained.
Second, it would be a window into the early Universe cosmic history. 
It would support the realization of an EMD phase,  during which deviations from spherical symmetry can significantly grow,  
and it would provide additional evidence for the presence of BSM physics. In our analysis we did not specify any underlying mechanism that realizes an EMD era but we remained agnostic and we let the reheating temperature to vary. 
A detection event  would also probe 
the primordial  spectrum of curvature perturbations which should 
not remain scale-invariant
within the entire momentum range. This last  would also have important implications for the inflationary model building directions.

 Our result has been produced by  a cold-collapse simulation which uses  Newtonian dynamics.  The cost of  neglecting relativistic effects likely introduces an error of order  $(v/c)^2$  which is typically negligible except when the system is near its gravitational radius. In this respect we also assume that the effect of the radiation back-reaction is unimportant, which is justified by the size of $dE/dt \ll 1$.  We presume that our results will not change if  the number of particles of our $N$-body simulations increases by several orders of magnitude, i.e.  beyond our computational capability.
We remark that the initial conditions for the $N$-body simulations are given by the   Zel'dovich solutions at the moment of the turnaround.
 This choice, though it might introduce  a systematic but presumably minor error in our results, it  makes possible a consistent  analysis  of a large collection of vastly  different initial configurations. 
Special attention was paid in the fitting procedure  for the components of the quadrupole tensor, in which our result has an increased sensitivity.
  We also comment that we considered a collisionless collapse,  
 neglecting  for simplicity any effect coming from decays and the subsequent reheating of the Universe.
 A next research step could be to improve upon  these issues. 
 
 In conclusion, our study  advances our understanding of certain aspects of the gravitational collapse in a (early) matter domination era.
Ultimately, it is exciting that  the sensitivity and the variety of  GW detectors increases and we can expect that predicted gravitational  signals from primordial Universe can be tested
probing  the very early Universe and its evolution from the Big Bang till today.

\section*{Acknowledgments}
ID would like to thank Maria Chira for her assistance
with the $N$-body simulations.
ID acknowledges support 
 by the Cyprus Research and
Innovation Foundation grant EXCELLENCE/0421/0362.


\end{document}